\newlength\smallfigwidth
\newlength\tinyfigwidth
\documentclass[aps,prb,twocolumn,floatfix,amsmath,amssymb,showpacs,showkeys]{revtex4}

\newcommand{\be}{\begin{equation}}
\newcommand{\ee}{\end{equation}}
\newcommand{\ba}{\begin{align}}
\newcommand{\ea}{\end{align}}
\newcommand{\bn}{\begin{eqnarray}}
\newcommand{\en}{\end{eqnarray}}

\usepackage{graphicx}
\usepackage{amsmath}
\usepackage{hyperref}

\begin{document}


\title{Dynamics and hysteresis in square lattice artificial spin ice}
\author{G.\ M.\  Wysin}
\email{wysin@phys.ksu.edu}
\homepage{http://www.phys.ksu.edu/personal/wysin}
\affiliation{Department of Physics, Kansas State University, Manhattan, KS 66506-2601}

\author{W. A. Moura-Melo}
\email{winder@ufv.br}
\homepage{https://sites.google.com/site/wamouramelo/}
\affiliation{Departamento de F\'{i}sica, Universidade Federal de Vi\c{c}osa, 36570-000 - Vi\c{c}osa - Minas Gerais - Brazil.}
\author{L. A. S. M\'{o}l}
\email{lucasmol@ufv.br}
\affiliation{Departamento de F\'{i}sica, Universidade Federal de Vi\c{c}osa, 36570-000 - Vi\c{c}osa - Minas Gerais - Brazil.}
\author{A. R. Pereira}
\email{apereira@ufv.br.}
\homepage{https://sites.google.com/site/quantumafra/home}
\affiliation{Departamento de F\'{i}sica, Universidade Federal de Vi\c{c}osa, 36570-000 - Vi\c{c}osa - Minas Gerais - Brazil.}

\date{February 21, 2013}
\begin{abstract}
{Dynamical effects under geometrical frustration are considered in a model for artificial spin 
ice on a square lattice in two dimensions.  
Each island of the spin ice has a three-component Heisenberg-like dipole moment subject to shape 
anisotropies that influence its direction.  The model has real dynamics, including rotation of the 
magnetic degrees of freedom, going beyond the Ising-type models of spin ice.  The dynamics is studied 
using a Langevin equation solved via a second order Heun algorithm.  Thermodynamic properties such as the 
specific heat are presented for different couplings.  A peak in specific heat is related to a type of 
melting-like phase transition present in the model.  Hysteresis in an applied magnetic field is calculated 
for model parameters where the system is able to reach thermodynamic equilibrium.}
\end{abstract}
\maketitle

\section{Introduction: Square spin ice, frustration, dynamics}
\label{intro}
Artificial spin ices are systems in two dimensions that mimic the usual three-dimensional spin ice materials that 
exhibit geometrical frustration effects: not all the pairwise spin interactions can be satisfied 
simultaneously \cite{Anderson56,Balents10,Moessner06,Castelnovo08,Ryzhkin05}. The name spin ice comes from 
the fact that lowest energy states obey the ice rule. For a square lattice, at each vertex where four spins meet, 
two point inward while two point outward.  Artificial spin ice compounds are built from 
magnetic nanoislands (typically, permalloy) which can be organized in different geometries
where the frustration is manifested\cite{Wang06,Li10,Ladak10,Mengotti11,Mol12,Reichhardt12,Budrikis12,Silva12b,Nascimento12}. 

Here, our focus is on artificial square lattice spin ice, first fashioned and studied by 
Wang \textit{et al.} \cite{Wang06} in $2006$. Artificial square ice consists of magnetic nanoislands 
(with a shape that looks like a ``cigar'') arranged as shown in Ref.\ \cite{Wang06} and 
here in Fig.\ \ref{spinD16T3h0}. Each nanoisland contains a net magnetic moment that tends to point along 
its long axis. 
When the interactions between neighboring islands are increased, the system increasingly 
fills with vertices that obey the two-in/two-out ice rule. Despite this, the predicted ground state of square 
ice was not observed experimentally until the work by Morgan \textit{et al.} \cite{Morgan11} in $2011$. Using 
magnetic force microscopy, those authors observed large regions of their samples which were able to 
adopt the square ice ground state. Morgan \textit{et al.} also observed the predicted excitations above the ground 
state, which resemble magnetic monopoles connected by energetic strings \cite{Morgan11,Mol09,Mol10,Moller09} (similar 
to Nambu monopoles \cite{Nambu74}). These elementary excitations are different from those of natural three-dimensional 
spin ices, which are magnetic monopoles connected by observable but non-energetic strings 
\cite{Castelnovo08,Ryzhkin05}. Therefore, these artificial compounds have attracted great interest in recent years.

Thermal activation of the island's magnetic moment (spin) configurations is very weak or nonexistent,
particularly with square lattice ice. Lack of 
thermalization is an important topic for experimental artificial spin ices and it can be partially alleviated 
by applying varying external magnetic fields \cite{Li10,Nisoli10}. Moreover, reductions in island volume and 
magnetic moment through state-of-the-art nanofabrication can bring energy scales closer to room temperature, 
leading to thermally driven slow dynamics. Alternatively, the use of materials with an ordering temperature near 
room temperature seems to be another important possibility. By using such a material, a recent experimental 
work on a square lattice in an external magnetic field confirms a dynamical ``pre-melting'' of the artificial spin 
ice structure at a temperature well below the intrinsic ordering temperature of the island material, creating a 
spin ice array that has real thermal dynamics of its artificial spins over an extended temperature range 
\cite{Kapaklis12}. Better understanding of these compounds may even come from
colloidal systems, which have an advantage over the usual magnetic arrays because thermal activation of 
the effective spin degrees of freedom is possible \cite{Reichhardt12}.  So, a more detailed analysis of the 
effects of thermal fluctuations and the spin dynamics in a two-dimensional spin ice material should be of 
great interest for a better understanding of these interesting frustrated systems.

Using an Ising model for the magnetic moments of the nanoislands, thermal effects in artificial square ice were 
studied recently by some of us \cite{Silva12} with Monte Carlo simulation. The focus was 
to examine the roles of elementary excitations in the thermodynamic properties of these systems. We found that the 
specific heat and average separation between monopoles with opposite charges exhibit a sharp peak and a local maximum, 
respectively, at the same temperature \cite{Silva12}, $T_{p}\approx 7.2D/k_{B}$, where $D$ is the strength of the 
dipolar interactions and $k_{B}$ is Boltzmann's constant. 
The Ising behavior of the islands seems to be realistic for the typical artificial magnetic ices made of 
permalloy (Py).  However, an Ising-type model does not display real time dynamics and it may also incorrectly 
estimate the degree to which energy barriers in dipolar reversal prevent thermalization. In this article, we study 
possibilities beyond the Ising behavior for the nanoislands. Our attention shifts to magnetic ices with real 
dynamics and the extra features that such dynamics may produce.  The internal structure 
and shape of the magnetic nanoislands is taken into account, assuming they are small enough to remain 
quasi-single-domain during reversal (with nearly coherent dipole rotation). 

The theoretical study of the net magnetic moment (with more degrees of freedom) of individual magnetic 
nanoislands with different sizes and shapes is the initial point.  
We presented a detailed study of non-Ising behavior for individual islands in Ref.\ \cite{Wysin+12}, which
also verified the coherent rotation of the dipole moment at small island sizes with high aspect ratios. 
Based on that, we assume that a nanoisland's spin is free to point in any possible direction, 
but with strong shape anisotropy energies that favor preferred directions. Then, 
differently from previous articles published on this topic, which consider only the 
dipolar interactions among the islands, here, these additional anisotropy terms are included 
in the Hamiltonian.
Thus, using a Langevin dynamics approach we have studied different models for
possible artificial square spin ices. Our results indicate that systems exhibiting real dynamics are feasible, in such a way that
their ground states could be achieved. On the other hand, for ordinary realizations with Py islands the system is not thermally driven to
its ground state indicating a possible dynamical bottleneck, abscent in systems with real dynamics. 

The article is organized as follows: In Section \ref{model}, the model is explained in detail. We define 
two order parameters to identify whether the ground state can be accessed at low temperature.  Three models 
(denoted by A,B,C) with different lattice and island parameters are studied to see the possibility of 
thermalized spin ice dynamics.  In Section \ref{thermal}, some thermal equilibrium properties of the models 
A,B,C are calculated.  In Section \ref{hyster}, we present  some hysteresis calculations and, finally, 
some discussions and conclusions are given in Section \ref{conclude}. 

\section{The model system}
\label{model}
The open square ice system with $N_c=L_1 \times L_2$ unit cells can be set up as follows. One can define the sites
of a square lattice in the usual way: the $k^{\rm th}$ lattice site (a monopole charge center or vertex) is at 
a point $\vec{r}_k=(x_k,y_k)$,  where $x_k=m_k a$ and $y_k=n_k a$ are integer multiples of the chosen lattice 
constant $a$. The points are chosen to fit inside the desired $L_1 \times L_2$ area.  For each unit cell of size
$a \times a$ there are two nano-islands that act as a two-atom basis, with the locations,
\bn
\label{sites}
\vec{r}_{k1} &=& (m_k+\tfrac{1}{2},n_k)a, \nonumber \\
\vec{r}_{k2} &=& (m_k,n_k+\tfrac{1}{2})a,
\en
where the ``1'' and ``2'' refer to the sublattices.
The nano-island on the 1$^{\rm st}$ sublattice has its long axis in the $x$ direction; the other nano-island,
on the 2$^{\rm nd}$ sublattice, has its long axis in the $y$ direction.  There are $N=2N_c$ islands in the
whole system.

A 3D vector magnetic moment $\vec\mu_{i}, i=1,2,3...N$ is associated with each island, whose defined
center position is some $\vec{r}_i$. Each $\vec{r}_i$ is selected from the set of $\vec{r}_{k\sigma}$,
where $\sigma=1,2$ denotes the sublattice. We use indeces $k,l$ for locations of the unit cells, and
indeces $i,j$ for locations of individual islands or their dipoles.

\begin{figure}
\includegraphics[width=1.2\smallfigwidth,angle=0]{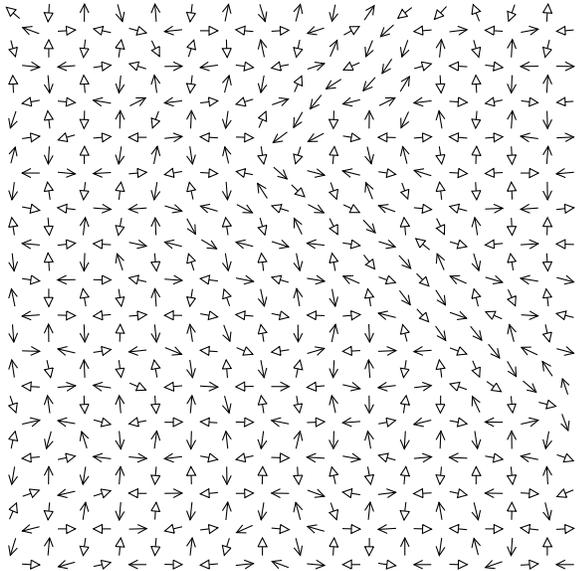}
\caption{\label{spinD16T3h0} A $16\times 16$ model system with $d=k_1=k_3=0.1$, in a metastable state
at temperature $k_B T/\varepsilon=0.025$, from a hysteresis scan (this is a state at $h_{\rm ext}=0$).
Most of the system is locally close to the $Z=+1$ ground state.  The upper right hand corner is locally
near the $Z=-1$ ground state, and there is a bent domain wall connecting the two regions. For
interior charge sites (junction points of four islands), there happens to be no discrete monopole
charges present: all $q_k=0$ and the discrete $\rho_m=0$.}
\end{figure}

The dipoles are assumed to have fixed magnitude $\mu$, while their direction is represented by a unit vector
$\hat\mu_i$.  The magnetic moments interact via long-range dipole forces, and are also affected by two forms
of local shape anisotropy.  First, there is a uniaxial anisotropy that impedes free rotation in the $xy$-plane,
associated with some energy constant $K_1$, and oriented along $x$ for the first sublattice and along
$y$ for the second.  Dependent on its sublattice, each moment has an axis $\hat{u}_i$ (equal to $\hat{x}$ 
or $\hat{y}$) for this anisotropy, see Fig.\ \ref{spinD16T3h0}.  Second, because the nano-islands are 
thin in the $z$-direction, the $z$-direction is a hard axis, and there is a hard-axis anisotropy whose 
energy scale is determined by a constant $K_3$, the same for all the islands.   The Hamiltonian is then
\bn
{\cal H} &=& -\frac{\mu_0}{4\pi} \frac{\mu^2}{a^3}
\sum_{i>j} \frac{ \left[ 3(\hat{\mu}_i\cdot \hat{r}_{ij})(\hat{\mu}_j\cdot\hat{r}_{ij})
                                -\hat{\mu}_i\cdot \hat{\mu}_j \right]}
{\left( {r}_{ij} / a\right)^3} \\
&+&  \sum_{i} \left\{ K_1[1-(\hat\mu_{i}\cdot\hat{u}_i)^2] + K_3 (\hat\mu_{i}\cdot \hat{z})^2
-\vec\mu_i \cdot \vec{B}_{\rm ext} \right\}
\nonumber
\en
Here $\mu_0$ is the magnetic permeability of space, and $\hat{r}_{ij}$ is the unit vector pointing
from the position of $\vec\mu_j$  towards the position of $\vec\mu_i$.  The first sum is the dipole-dipole
interactions, the second sum contains the anisotropy energies and an applied external magnetic induction
$\vec{B}_{\rm ext}=\mu_0 \vec{H}_{\rm ext}$.  A constant is included in the $K_1$ anisotropy energy so
that that energy is zero when a dipole points along its local anisotropy axis $\hat{u}_i$.
Note that if a dipole moves in the $xy$ plane, it only pays the cost of the $K_1$ anisotropy term, 
but motion up out of the $xy$ plane (say, in the $xz$ plane) involves an energy proportional to the 
sum of both anisotropies, $K_1+K_3$.

The motion out of the $xy$ plane is also impeded by the dipolar interactions.  With the dipole pair
distances scaled by the lattice constant, the effective strength of nearest neighbor dipolar interactions
is determined by the dipole energy factor,
\be
D = \frac{\mu_0}{4\pi} \frac{\mu^2}{a^3}.
\ee
Depending on the island geometry, which is discussed further below, the anisotropy constants $K_1$ and $K_3$
would typically be of similar order of magnitude.  Thus, there are three important energy scales:  dipolar
energy, anisotropy energy, and the thermal energy $k_B T$.   The anisotropy constants are proportional to
the volume $V$ of the islands, as
is $\mu=M_s V$, where $M_s$ is the saturation magnetization of the magnetic material.  But then, this
dipolar constant $D$ increases as the squared island volume.  Thus, changing the island
size and spacing $a$ can be used to adjust these energy scales in relation to each other.  Typically, the
interesting case must have the thermal energy less than both the effective dipolar energy (per site) and
anisotropy energy.   But note, the effective dipolar energy can be quite a lot larger than that indicated
by $D$, which only measures the energy in a nearest neighbor pair.   When the dipolar interactions are
summed,  the net dipolar energy per island could be much larger than $D$.

\subsection{Spin-ice ground state and order parameters}
For the square lattice spin-ice, the ground state is two-fold degenerate, and involves alternating
dipoles on each of the two sublattices.  The ground state fully satisfies the two-in/two-out rule in
each monopole charge cell (junction of four islands at the site $\vec{r}_k$ of each unit cell).
The unit cell positions are expressed $\vec{r}_{k}=(m_k,n_k)a$, where $a$ is the lattice constant and
$m_k$ and $n_k$ are integers.  Then one of the ground states can be constructed by setting the dipole
directions as:
\bn
\label{GS}
\hat\mu_{k1}^{\rm GS} &\equiv& \hat\mu_1^{\rm GS}(\vec{r}_{k}) =  +(-1)^{m_k+n_k} ~ \hat{x}, \nonumber \\
\hat\mu_{k2}^{\rm GS} &\equiv& \hat\mu_2^{\rm GS}(\vec{r}_{k}) =  -(-1)^{m_k+n_k} ~ \hat{y}.
\en
This formula is arranged so that at a chosen unit cell at position $\vec{r}_{k}$, the dipole on one
sublattice points inward and the dipole on the other sublattice points outward, thereby globally enforcing
the two-in/two-out rule.  By reversing the sign on all the dipoles, the other ground state is obtained.

With the ground state determined, we can construct a measure of the proximity (in phase space) of any arbitrary
state to one of the ground states.   This order parameter $Z$ is simply the overlap with this ground state:
\be
Z \equiv \langle \psi^{\rm GS} \vert \psi \rangle =
\frac{1}{2N_c} \sum_{k=1}^{N_c} \sum_{\sigma=1}^{2} \hat\mu_{k\sigma}^{\rm GS}\cdot \hat\mu_{k\sigma} .
\ee
The index $\sigma$ labels the sublattice.  If the system happens to be found in the ground state
defined in Eq.\ (\ref{GS}), then $Z=1$; if the system is in the inverted ground state, then $Z=-1$.
Thus it is possible to show that the range of $Z$ is from -1 to +1.  This order parameter is useful
for indicating the degree of thermodynamic excitation in the system, by the deviation of $|Z|$ from
unity.  Further, its sign then gives an indication of processes which involve the transformation from
one ground state to the other.   Indeed, considered even as a local variable (calculating only near a
single charge cell), we can track when the system has different regions close to either of the ground
states, possibly with regions separated by domain walls.  Fig.\ \ref{spinD16T3h0} shows an example,
where the system has a region near one of the ground states, with $Z\approx +1$, separated by a
domain wall from another region that is near the other ground state, with $Z\approx -1$.  The net
averaged value of $Z$ for the entire system, however, acquires an intermediate value, $Z\approx 0.521$,
indicating considerable separation from a uniform ground state.

The other obvious order parameter to be measured is the areal density of monopole charges, $\rho_m$.
We make a simple discrete definition, to connect to Ising spin ice models, and, a more generalized
continuous definition that accounts for the greater freedom of the continuous dipoles in the model
described here.  The discrete definition of a monopole charge involves counting the net number
of dipoles that point outward at a chosen charge site $\vec{r}_{k}$, and dividing that result by two.
There are four dipoles $\vec\mu_{i_k}, i_k=1,2,3,4$, surrounding any charge cell center $\vec{r}_{k}$.
Then the possible monopole charge values are $q_k = 0, \pm 1, \pm 2$; the double charges, $q_k=\pm 2$, 
may typically be of low probability but contribute doubly to the charge density.   For the discrete 
charge definition, whether a dipole points \textit{outward} or \textit{inward} is determined with a 
Heaviside step function $H(x)$:
\be
\label{qk_discrete}
q_k = \frac{1}{2} \sum_{i_k=1}^{4} \left[ 2H(\hat\mu_{i_k} \cdot \hat{v}_{i_k}) -1 \right] .
\ee
The unit vectors $\hat{v}_{i_k}, i_k=1,2,3,4,$ point outward from charge site $\vec{r}_{k}$ to each
of the four nearest islands.

Because this discrete definition can show sudden change when a dipole rotates 90$^{\circ}$ from
the radially outward direction, we also considered a continuous definition.  In the continuous definition,
the step function is removed, and only a scalar product is needed,
\be
\label{qk_contin}
q_k =  \frac{1}{2} \sum_{i_k=1}^{4}  \hat\mu_{i_k} \cdot \hat{v}_{i_k} .
\ee
In contrast to the discrete definition, this charge definition varies continuously from $q_k=-2$ to $q_k=+2$.
One can also note that for either the discrete or the continuous definition, total monopole charge is
conserved.  A positive contribution produced by some dipole at one charge cell is accompanied by
an equal negative contribution at a neighboring charge cell (each dipole contributes to two charge cells).
Then, the total algebraic monopole charge in the system takes the conserved value, zero.

In order to get a measure of the monopole charges present, regardless of their sign, we define a density
for the system as a whole, by applying absolute value.  Thus, the ``monopole density'' measured in the
simulations here is defined as
\be
\label{rho}
\rho_m = \langle \vert q_k \vert \rangle = \frac{1}{N_c} \sum_{k=1}^{N_c} \vert q_{k} \vert
\ee
This is averaged over charge cells.  By using absolute value, the definition does not allow
the cancellation of charges of opposite signs. Note that in either of the ground states, there are no charges
at any sites, and $\rho_m=0$.  Charges appear as the system moves away from the ground state.  
(This is true for the spin ice on the square lattice, but not on the Kagom\'e lattice, whose ground state 
contains charges, due to there being three dipoles for each charge cell.)
Thus, this is another measure of excitation in the system.  

At very high temperature, the individual dipoles can point freely in all directions.  In this high-entropy
limit, the value of $\rho_m$ from both definitions can be determined.  For the discrete definition, each
of the four dipoles in a vertex are in or out with equal probabilities.  Of the 16 possible states, there 
are 6 with $q_k=0$, 8 with $q_k=\pm 1$ and 2 with $q_k=\pm 2$.  The average charge
per vertex, including single and double charges,  is
\be
\rho_m = \langle \vert q_k \vert \rangle = (6 \times 0 + 8 \times 1 + 2 \times 2)/16 = 3/4. 
\ee
For the continuous definition, there is a corresponding expression from averaging over the
sum of projections $x_{i_k} \equiv \hat\mu_{i_k} \cdot \hat{v}_{i_k}$ of dipoles on their local axes, 
see (\ref{qk_contin}).  With each $x_{i_k}$ ranging from $-1$ to $+1$, we have the average 
in an arbitrary cell
\be
\rho_m = 
\frac{\int dx_1 ~ dx_2 ~ dx_3 ~ dx_4 ~ \frac{1}{2}\vert x_1 + x_2 + x_3 +x_4 \vert}
{\int dx_1 ~ dx_2 ~ dx_3 ~ dx_4 ~ 1} =  7/15.
\ee
We note that while these should be the limits in the state of greatest disorder, they are
not upper limits.  One can observe that, for example, by taking the ground state configuration
and reversing the dipoles only on one sublattice, a state will be obtained that has a doubly-charged
monopole in every cell.  That state would have $\rho_m=2$ by both definitions.  Thus, the whole range 
$0\le \rho_m \le 2$ is allowed.

\subsection{The undamped dynamics}
The zero-temperature, undamped dynamics of each magnetic dipole is determined by a torque equation,
\be
\frac{d\vec\mu_i}{dt} = \gamma \vec\mu_i \times \vec{B}_i
\ee
where $\vec{B}_i$ is the local magnetic induction acting on the $i^{\rm th}$ dipole, and $\gamma$
is the electronic gyromagnetic ratio.  The local magnetic induction is derived from the Hamiltonian by
assuming an energy $-\vec\mu_i\cdot \vec{B}_i$ for each dipole, i.e.,
\bn
\label{Bi}
\vec{B}_i &=& -\frac{\delta {\cal H}}{\delta \vec\mu_i} = -\frac{1}{\mu}\frac{\delta {\cal H}}{\delta \hat\mu_i}
= \frac{D}{\mu} \sum_{j\ne i} \frac{3 (\hat\mu_j\cdot\hat{r}_{ij}) \hat{r}_{ij}-\hat\mu_j}{(r_{ij}/a)^3}
\nonumber \\
&+& 2\frac{K_1}{\mu} (\hat\mu_i\cdot \hat{u}_i) \hat{u}_i
- 2\frac{K_3}{\mu} (\hat\mu_i\cdot \hat{z}) \hat{z} + \vec{B}_{\rm ext}.
\en
It will be convenient to choose some standard units for the time, the applied field, and so on, to
simplify and scale the numerical calculations.  The dipole terms are simplified by selection of the
lattice constant $a$ as the unit of length.   A natural unit to measure field $\vec{H}_{\rm ext}$
is the saturation magnetization $M_s$ from which the particles are made.  For example, for permalloy,
with $M_s = 860$ kA/m, this unit as a magnetic induction is close to one tesla:
$\mu_0 M_s \approx 1.08$ T.  Using this quantity to scale the magnetic field and hence the magnetic
induction defines their dimensionless field,
\be
\vec{h}_{\rm ext} = \frac{\vec{H}_{\rm ext}}{M_s} = \frac{\vec{B}_{\rm ext}}{\mu_0 M_s}.
\ee
When $\vec{h}_{\rm ext}$ approaches 1.0 the applied field should have a strong tendency to
saturate the magnetization of the system (if the dipolar interactions do not impede that).
This then indicates how to scale the dipole and anisotropy fields, i.e., by writing the
dimensionless local magnetic fields from (\ref{Bi}),
\bn
\label{hi}
\vec{h}_i &=& \frac{\vec{B}_i}{\mu_0 M_s}
= \frac{D}{\mu \mu_0 M_s} \sum_{j\ne i} \frac{3 (\hat\mu_j\cdot\hat{r}_{ij}) \hat{r}_{ij}-\hat\mu_j}{(r_{ij}/a)^3}
 \\
&+& 2\frac{K_1}{\mu \mu_0 M_s} (\hat\mu_i\cdot \hat{u}_i) \hat{u}_i
- 2\frac{K_3}{\mu \mu_0 M_s} (\hat\mu_i\cdot \hat{z}) \hat{z} + \vec{h}_{\rm ext}.
\nonumber
\en
This involves dimensionless coupling constants that indicate the relative
strength of each contribution,
\bn
\label{dk1k3}
d &=& \frac{D}{\mu\mu_0 M_s} = \frac{\mu}{4\pi a^3 M_s},
\\
k_1 &=& \frac{K_1}{\mu\mu_0 M_s}, \quad
k_3 = \frac{K_3}{\mu\mu_0 M_s}.
\en
These definitions involve the different energy scales divided by an energy unit,
\be
 \varepsilon \equiv \mu_0\mu M_s,
\ee
that depends on the size of the magnetic islands.
The dimensionless dipole parameter $d$ can be seen to be proportional to the volume fraction of the
system occupied by magnetic islands, since $\mu = M_s V$ for each island.  Obviously a higher packing of
magnetic material into the lattice leads to stronger dipolar effects, and $d$ indicates their effective
strength.

The dynamic equation can be scaled in the same way, so that the dimensionless field appears on
the RHS. Then the dynamics for the unit vector dipoles is described using a rescaled time $\tau$,
\be
\frac{d\hat\mu_i}{d\tau} = \hat\mu_i \times \vec{h}_i, \quad
\tau = \gamma \mu_0 M_s\, t.
\ee
With the above scaling of the fields, the unit of time is $(\gamma\mu_0 M_s)^{-1}$.
For the case of permalloy and using the gyromagnetic ratio as
$\gamma = e/m_e \approx 1.76 \times 10^{11}$ T$^{-1}$ s$^{-1}$,
this unit is $(\gamma\mu_0 M_s)^{-1} \approx 5.26 $ ps.

\subsection{Island geometry and energetics}
The shape anisotropy constants $K_1$ and $K_3$ can be estimated based on the magnetic properties
for permalloy (or other material) and micromagnetics simulations for the choice of island geometries
and island volume $V$.  We consider thin elliptical islands.  Here $L_x$ denotes
the major-diameter of the ellipse and $L_y$ is the minor-diameter, while $L_z$ is the height of
the island or its thickness.  The semi-major axis is $A=L_x/2$, the semi-minor axis is $B=L_y/2$.
It is well-known that an elliptically shaped magnetic particle will have anisotropy
\cite{Wei03} within the plane of the island.
In Ref.\ \cite{Wysin+12},  the anisotropy constants (as energies per unit volume) were estimated based
on a calculational approach for thin elliptical islands, for a range of thicknesses $L_z\ll L_x$,
characterized by an aspect ratio $g_3=L_x/L_z$,  and various lateral aspects ratio $g_1=L_x/L_y$.
Here we consider some different sizes and shapes for the islands  and discuss the expectations for their 
dynamics and the relative importance of the different energy scales when placed in a square spin-ice array.

\textbf{Model A}. In Wang \textit{et al.} \cite{Wang06},  experiments on square spin-ice were carried
out for (quasi-rectangular) particles with dimensions 220 nm $\times$ 80 nm $\times$ 25 nm, where the
last number is the vertical thickness.  In those experiments, the particle sizes were kept fixed,
but different lattice parameters $a$ from 320 nm to 880 nm were used.   In this first model,
we use these numbers to describe elliptical particles: $L_x=220$ nm, $L_y=80$ nm, $L_z=25$ nm.
Then the particle volume is $V=\pi AB L_z = 3.46 \times 10^5$ nm$^3$, and using a saturation
magnetization $M_s = 860 $ kA/m for Py, the magnetic dipole moment per particle is
$\mu= 2.97 \times 10^{-16}$ A$\cdot$m$^2$, the equivalent of about $3.2\times 10^7$ Bohr magneton.
For its aspect ratio parameters $g_1=2.75$ and $g_3=8.8$, the anisotropy energy densities can be found
by interpolation of the simulation results in Ref. \cite{Wysin+12}, as $K_1/V=0.0064$ $A_{\rm ex}$/nm$^2$ and
$K_3/V = 0.0143$ $A_{\rm ex}$/nm$^2$, where $A_{\rm ex} \approx 13$ pJ/m is the exchange stiffness for Py.
The easy-axis anisotropy is then $K_1 = 2.9 \times 10^{-17}$ joules, while the
hard-axis anisotropy is estimated as $K_3 = 6.4 \times 10^{-17}$ J.  These are considerably
larger than room-temperature (300 K) thermal energy  $k_B T\approx 4.1 \times 10^{-21}$ J, as needed
for stable magnetic moments.  The energy unit is $\varepsilon=\mu_0\mu M_s=3.21\times 10^{-16}$ J.
Then the dimensionless anisotropies are $k_1 = K_1/\varepsilon=0.0897$, $k_3=K_3/\varepsilon=0.200$ .
The scaled thermal energy at room temperature is ${\cal T} \equiv k_B T/\varepsilon=1.29\times 10^{-5}$,
an extremely small value.

The nearest neighbor dipolar energy scale might be estimated first at
lattice constant $a=880$ nm, for which it is $D=1.29 \times 10^{-20}$ J, about  2000 times
smaller than $K_1$.  If instead the lattice constant $a=320$ nm is used, this will scale up by a factor
of $(880/320)^3$, leading to $D=2.68 \times 10^{-19}$ J, or still 100 times smaller than $K_1$.
The dimensionless dipolar coupling for $a=320$ nm is $d=D/\varepsilon=8.35 \times 10^{-4}$.
Obviously, values of $k_1,k_3,$ and $d$ similar to these are needed to get
a spin-ice system, however, dynamics simulations are difficult with these parameters because
the anisotropy is so dominant and Ising-like.  Over the time scales that can be accessed in numerical
simulations, one would not expect to see much dynamical flipping of the island dipoles, except
in the presence of a strong applied magnetic field.  Thus it may be interesting instead to consider
some other particle sizes where the dynamics can be expected to be more active.

A thinner or smaller island will result in a smaller magnetic dipole moment $\mu$, which
leads linearly to weaker anisotropy, but quadratically to weaker dipolar energy.  Both energy scales
become closer to the thermal energy.   Thus we can try to change the particle size in such a way so
that room temperature thermal energy is closer to $K_1$ and perhaps even larger than $D$.


\textbf{Model B}. Here we consider smaller particles, with $L_x=40$ nm, $L_y=8.0$ nm, $L_z=4.0$ nm,
to try and get weaker anisotropy energy scales (for Py parameters).   The particle volume is
now only $V=1005$ nm$^3$, and the dipole moment is $\mu=8.64\times 10^{-19}$ A$\cdot$m$^2$.
At aspect ratio parameters $g_1=5.0, \quad g_3=10$,  the anisotropy energies are found to be
somewhat smaller: $K_1=1.38\times 10^{-19}$ J, and $K_3=1.12\times 10^{-19}$ J.  The energy unit,
though, is now also smaller: $\varepsilon=9.34\times 10^{-19}$ J, leading to dimensionless
couplings $k_1=0.148$ and $k_3=0.120$ . 
The smaller energy unit means that room temperature effects may be more accessible.
The scaled thermal energy at 300 K is increased: ${\cal T}=k_B T/\varepsilon=0.00443$ .
For a lattice with $a=80$ nm, we find $D=1.46\times 10^{-22}$ J, and $d=D/\varepsilon=1.56\times 10^{-4}$.

\vskip 0.1in
It is clear in the above examples that the strong Ising-like anisotropy for 
real spin-ice particles dominates over thermal energy at (and below) room temperature.  
That being the case, we find it interesting also to study a model with fictitious parameters,
which might be possible to achieve in other materials with different values of $M_s$, $K_1/V$, etc.

\textbf{Model C}.  Rather than assuming a particular particle size and using Py parameters,
suppose some particles are arranged so that $D=K_1=K_3=\tfrac{1}{10}\varepsilon$.  Obviously nature 
may not easily produce such a system with all equal energy scales, but it may be possible by
appropriate materials engineering.   We use a fraction of $\varepsilon$, which is required by
the definition of $d$, see Eq.\ (\ref{dk1k3}) (the volume fraction of dipoles on the lattice
cannot be more than unity).  The scaled energy parameters are all equal: $d=k_1=k_3=0.1$ .
A physical value of $\varepsilon$ is needed, based on values of $\mu$ and $M_s$ for some real
particles, to locate room temperature on the temperature scale.


\section{Thermal equilibrium properties}
\label{thermal}
Mostly the magnetic properties of spin-ice materials are investigated in an approximation of
zero temperature, because the fundamental interaction strengths of the anisotropy energies and
the dipolar energies are much greater than $k_B T$ at room temperature (Models A \& B).   Even so,
there could be  energetic thermal fluctuations in a magnetic system even at low temperatures, in
any situation where the magnetic fluctuations are large, such as near a reversal point in a hysteresis
loop.   This might lead to enhancement of specific heat in such a situation, and of course,
thermal rounding of the reversal paths in magnetization hysteresis loops.  Thus it could be
interesting to have some calculations of the energy, specific heat, and also of magnetic
susceptibilities in a situation of thermal equilibrium.

The time evolution from Langevin dynamics can be used to get thermal averages, as an
alternative to Monte Carlo calculations, that includes true dynamical effects.  Details of the 
Langevin simulation method, as
solved using a second order Heun integrator, with FFT for calculation of the dipole
fields, are given in the Appendices.  Provided the simulation time is long compared
to any physical relaxation time, a sequence of energy samples $E_n$ and total system
magnetization samples $\vec{M}_n$ can be averaged, and their fluctuations can be used to
estimate the specific heat and magnetic susceptibility.  Suppose there are $N_s$ samples
taken from the time evolution.   The average energy for all $N$ islands is estimated as
\be
\langle E \rangle = \frac{1}{N_s} \sum_n E_n
\ee
with a measurement error estimated from its standard deviation $\sigma_E$ and the number
of samples,
\be
\Delta E = \frac{\sigma_E}{\sqrt{N_s}}, \quad
\sigma_E^2 = \langle E^2 \rangle -\langle E \rangle^2.
\ee
The total heat capacity of the system is determined from the fluctuations in the energy,
\be
C_N = \beta^2 \langle \left( E-\langle E \rangle \right)^2 \rangle
= \beta^2 \sigma_E^2,
\ee
where $\beta = (k_B T)^{-1}$, and from that we obtain the specific heat per island, $C=C_N/N$.
The error in the heat capacity is calculated by finding the standard deviation of the quantity
$z\equiv \left( E-\langle E \rangle \right)^2$ from which $C_N$ was obtained, which is
found from the following averages,
\be
\label{sigmaz}
\sigma_z^2 = \langle E^4 \rangle -\langle E^2 \rangle^2 +4\langle E \rangle \left[
2\langle E^2 \rangle  \langle E \rangle -\langle E^3 \rangle -\langle E \rangle^3 \right].
\ee
Then the error in $C_N$ is
\be
\label{DC}
\Delta C_N = \beta^2 \frac{\sigma_z}{\sqrt{N_s}},
\ee
and the error in specific heat per island is $\Delta C = \Delta C_N/N$.  
Likewise, the susceptibility per island, $\chi_{xx}$, is found from
fluctuations in total magnetic moment of the system, $M_x=\sum_l \mu_l^x$,
\be
\chi_{xx} = \frac{\beta}{N} \left\langle \left( M_x-\langle M_x \rangle \right)^2 \right\rangle
= \frac{\beta}{N} \sigma_{M_x}^2,
\ee
and its error $\Delta \chi_{xx}$ comes from relations similar to (\ref{sigmaz}) and (\ref{DC}).

Due to the fluctuations caused by the temperature in the simulations, the calculations of
$C$ and $\chi$ are generally not as precise as those of $\langle E \rangle$ and $\langle \vec{M} \rangle$,
without making very long runs.   Especially as mentioned above, these calculations are difficult in
any physical situation where the magnetization is on the verge of reversal, where the fluctuations
are greatest.

The system was started in a random state, with the temperature initially
set at the highest value in the range of interest.  For a chosen temperature, data samples were taken
at some appropriate time interval that depends somewhat on the energy couplings.   For coupling
parameters $k_1,k_3$ on the order of 0.2 or less, and $d$ several orders smaller (Models A \& B),
a Heun time step $\Delta\tau=0.01$ was sufficient to insure proper energy conservation at
zero temperature.  Using this time step for finite temperature together with damping $\alpha=0.1$, we averaged
over $N_s=4000$ data samples separated by sampling time interval $\Delta\tau_s=10^3 \Delta\tau=10.0$ .
An initial time interval corresponding to 100 samples was allowed for relaxation before samples were taken.
Simulations would be left to run even longer than 4000 samples, if necessary, until the percent error in
the magnitude of the total system magnetization was found to be less than 0.1\%.   The final state at one
temperature was then used as the initial state for the next lower temperature in the calculation.

For model C, the dipolar coupling is much stronger, and this requires a smaller Heun time step,
$\Delta\tau=0.001$, to insure proper dynamics at finite temperature and energy conservation at zero
temperature.  Besides this, the calculation parameters for averages were the same as for models A \& B,
e.g., $N_s=4000$ and taking samples at sampling time interval $\Delta\tau_s = 10^3 \Delta \tau$,
while waiting for 0.1\% or better precision in the system magnetization.

\begin{figure}
\includegraphics[width=\smallfigwidth,angle=-90]{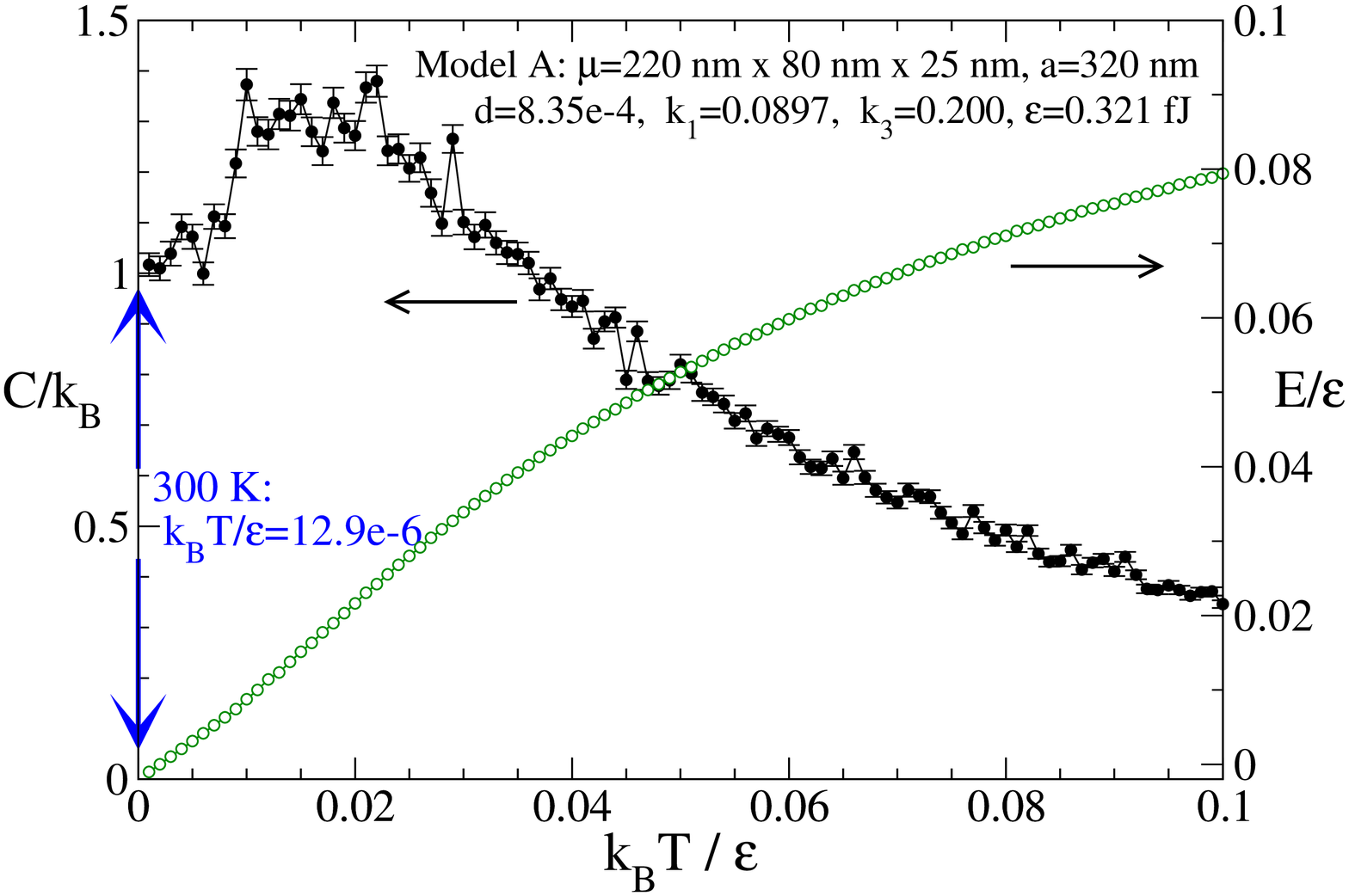}
\includegraphics[width=\smallfigwidth,angle=-90]{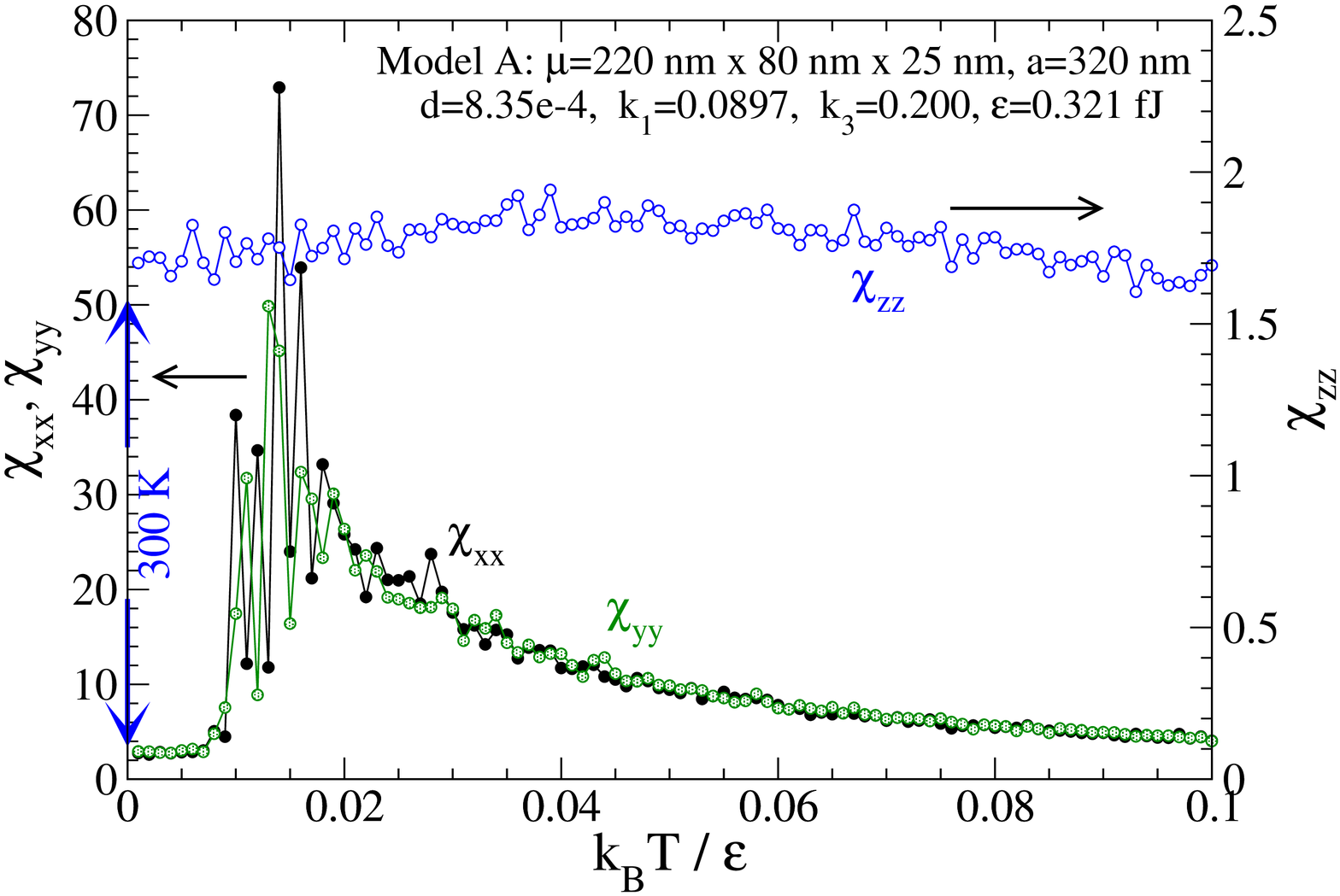}
\includegraphics[width=\smallfigwidth,angle=-90]{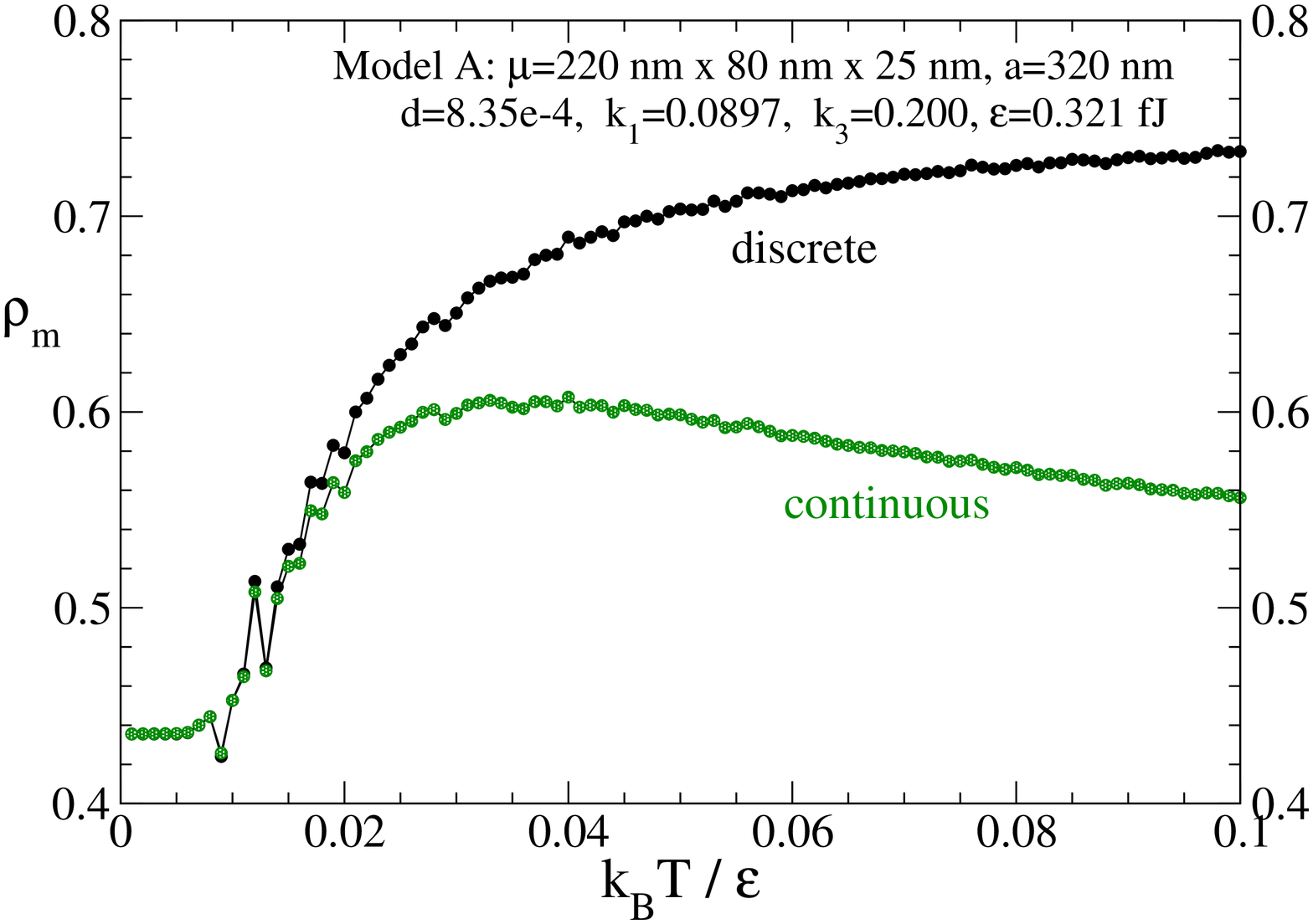}
\caption{\label{eckTA16} (Model A) For a $16\times 16$ grid of particles as used in Wang \textit{et al.}
with indicated parameters, (a) the internal energy and specific heat per site versus scaled temperature;
(b) the components of the magnetic susceptibility at zero external field;  (c) The monopole density,
Eq.\ (\ref{rho}) as determined from discrete and continuous charge definitions, Eqs.\
(\ref{qk_discrete}) and (\ref{qk_contin}).  The vertical arrows very near $k_B T/\varepsilon=0$ 
show room temperature: it is essentially unaccessible in this dynamics.}
\end{figure}


\begin{figure}
\includegraphics[width=\smallfigwidth,angle=-90]{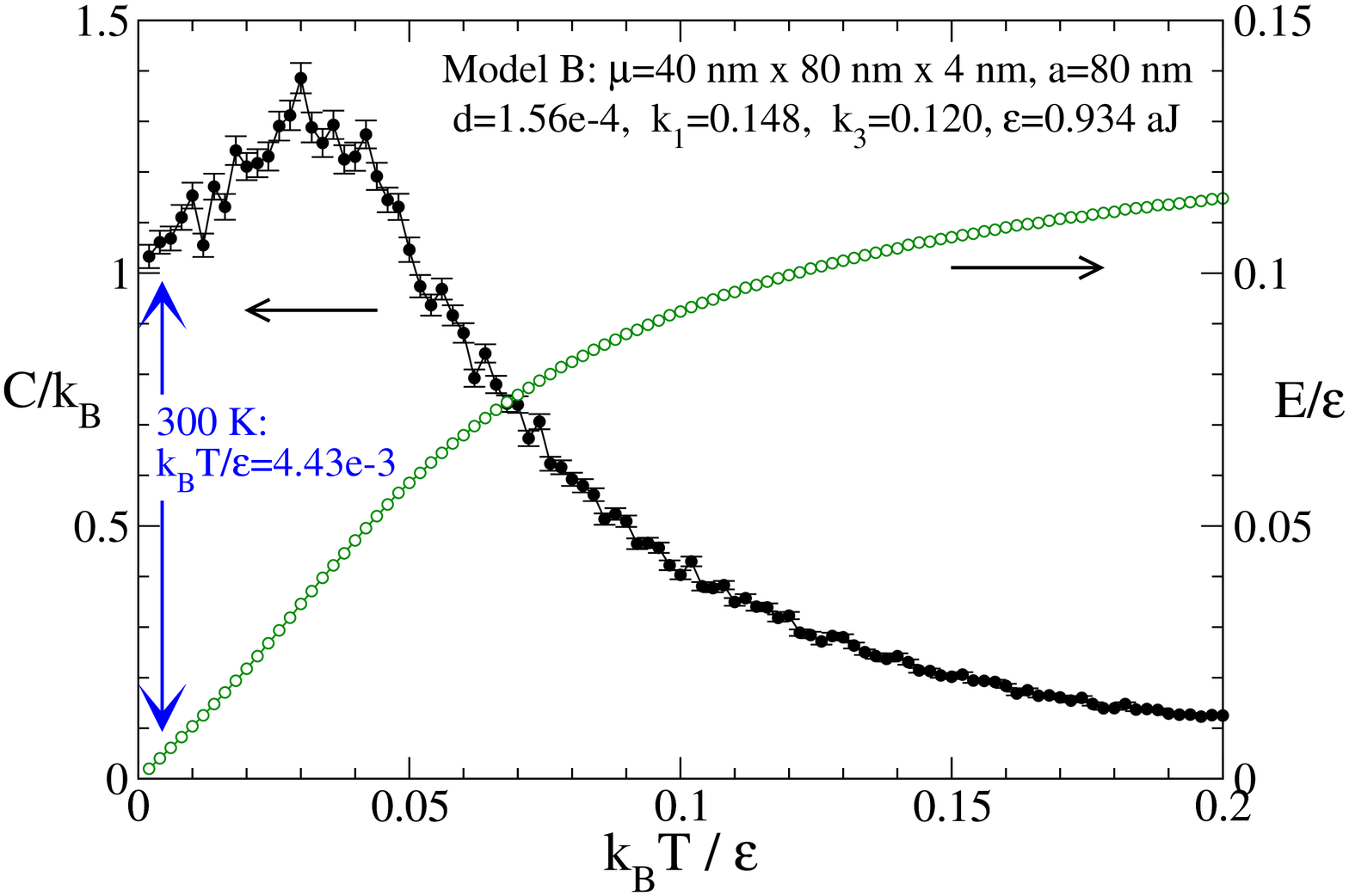}
\includegraphics[width=\smallfigwidth,angle=-90]{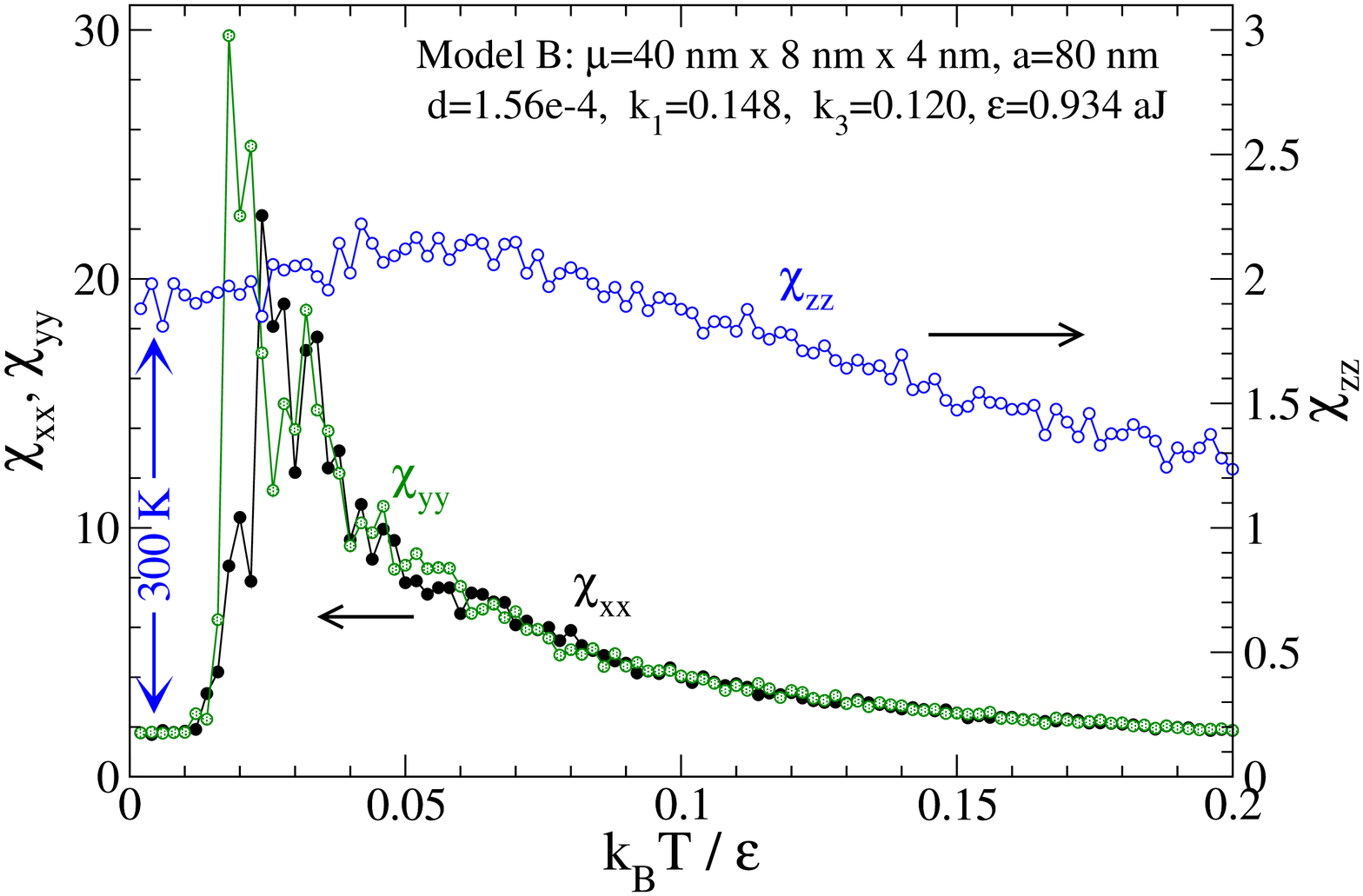}
\includegraphics[width=\smallfigwidth,angle=-90]{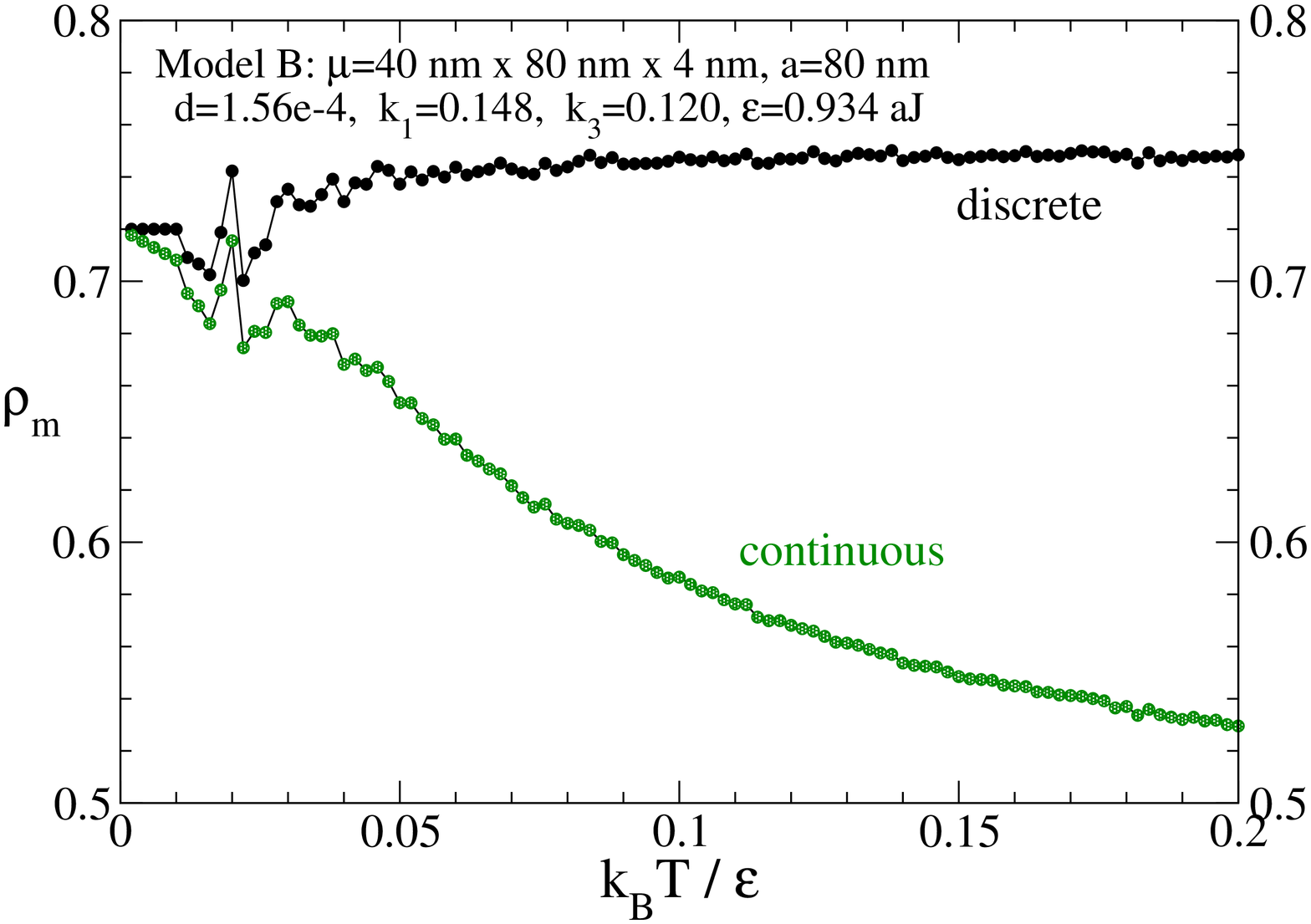}
\caption{\label{eckTC16} (Model B) For a $16\times 16$ grid of still smaller sized particles with indicated
parameters and an even lower energy scale, (a) the average internal energy per site and the specific
heat per site versus scaled temperature; (b) the components of the magnetic susceptibility at zero
external field; (c) the monopole charge density.}
\end{figure}

\begin{figure}
\includegraphics[width=\smallfigwidth,angle=-90]{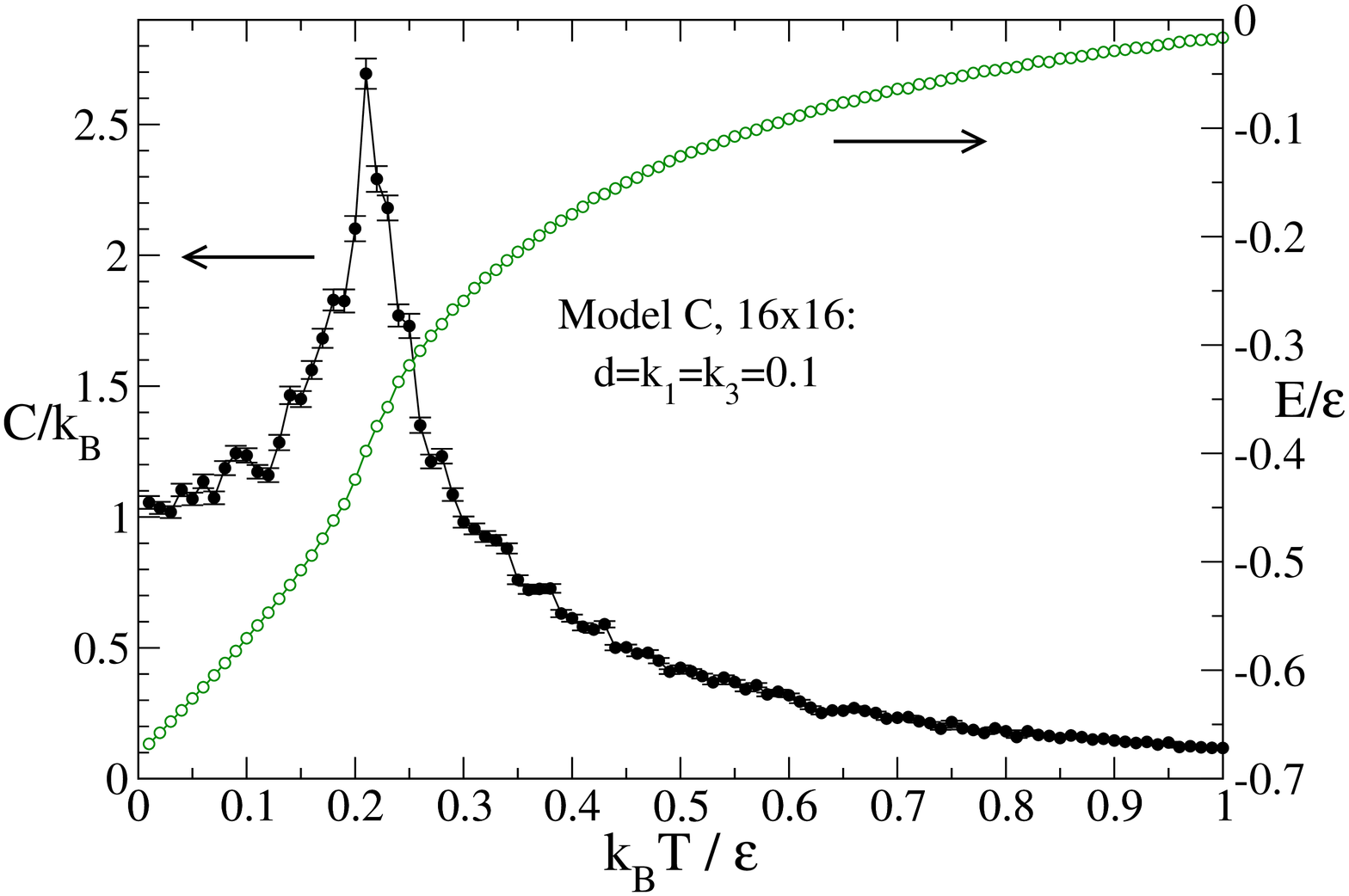}
\includegraphics[width=\smallfigwidth,angle=-90]{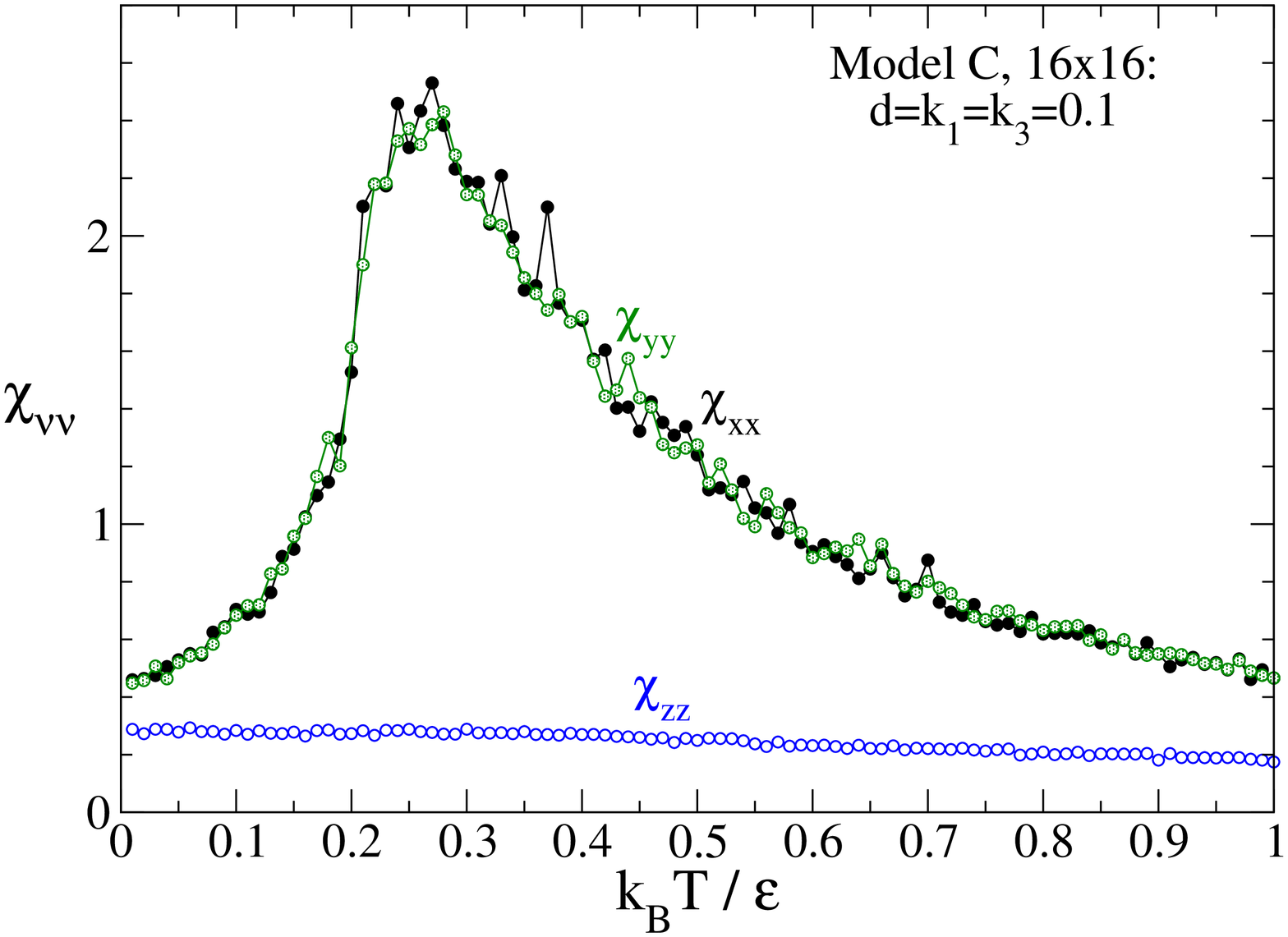}
\includegraphics[width=\smallfigwidth,angle=-90]{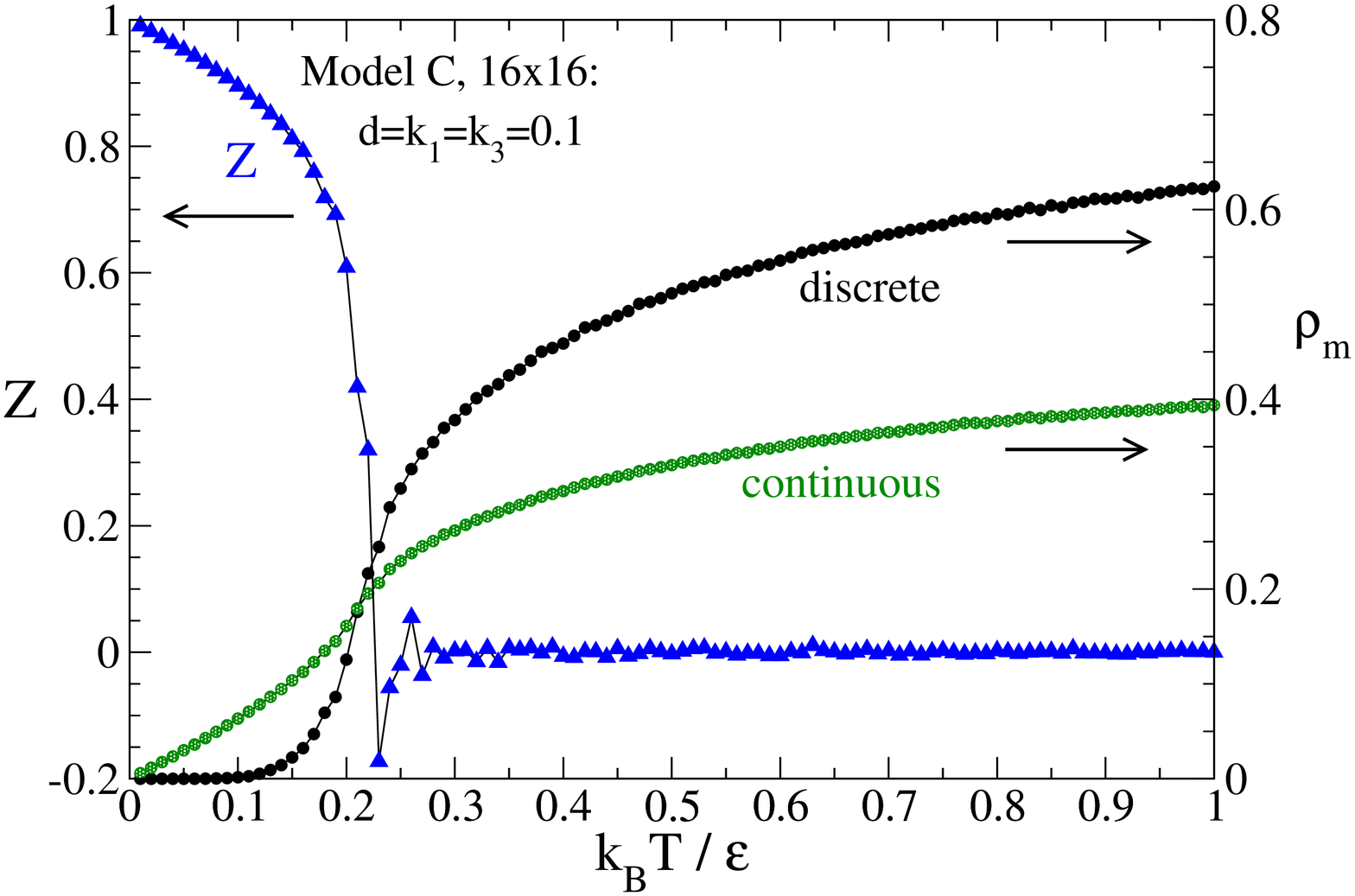}
\caption{\label{eckTD16} (Model C) For the model with $D=K_1=K_3=\tfrac{1}{10}\varepsilon$,
(a) the average internal energy per site (in energy units $\varepsilon$) and the resulting specific heat
per site versus scaled temperature; (b) the components of the magnetic susceptibility at zero external
field; (c) the monopole charge density together with the ground state overlap order parameter $Z$.}
\end{figure}

Some thermodynamic results  for the Wang \textit{et al.} particles (Model A) are
shown in Fig.\ \ref{eckTA16}, versus scaled temperature ${\cal T}=k_B T/\varepsilon$.  A $16\times 16$ grid of
cells was used ($N=2\times 16^2=512$).  As the dimensionless energy coupling
constants are small numbers, the only interesting effects are observed for ${\cal T}< 0.1$ .
Near ${\cal T}\approx 0.02$ there are peaks in specific heat and in the in-plane components
of magnetic susceptibility.   As mentioned earlier, though, these features would appear only
greatly above room temperature, which is marked with arrows.  At these ``high'' temperatures,
other modifications would take place first (besides magnetic effects) and the
model would not be applicable.  Note that the monopole charge density in Fig.\ \ref{eckTA16}c
does not go to zero at very low temperature here.  It does, however, make a transition to a
lower value.  This is an indication that the system did not find a state close to a ground state.
There is frozen-in disorder at lower temperatures.  It is also an indication that the time scale
for thermal relaxation to an equilibrium configuration was longer than the time interval used
for averaging. However, the specific heat per site does tend towards $C/k_B \rightarrow 1$ as $T\rightarrow 0$, 
consistent with the dipoles simply making small fluctuations around their local anisotropy axes 
(the long axes of the islands).  In contrast to this, an Ising model for this system would lead
to $C/k$ tending towards zero for low temperature.  Both the discrete and continuous definitions of 
$\rho_m$ exhibit similar behaviors, and they tend towards the expected high-temperature limits of 
3/4 and 7/15, respectively.  At very low temperatures they trend together and give a nearly 
identical limit as $T\rightarrow 0$.  The order parameter $Z$ (not shown) stayed close to zero 
for the whole temperature range shown.  That is further indication of the system staying far from 
a ground state, where it would have reached one of the values $\pm 1$.  The dipoles in this
limit are nearly aligned with the islands' long axes, however, with a frozen-in disorder,
not near a ground state.

Results for Model B's smaller particles are shown in Figure  
\ref{eckTC16}.  These confirm that for the typical square lattice spin-ice using Py as the material, 
the room-temperature thermodynamics is nearly the same as that at zero temperature.  There would be
some limiting specific heat $C\approx k_B$ and non-zero value for the in-plane susceptibility.
However, the monopole density has extreme difficulty to go to zero while scanning from high to
low temperature, although both the discrete and continuous definitions tend to the same value at very
low temperature.  Then, in fact, the dynamics is a low temperature dynamics in a disordered
non-ground state (and non-equilibrium) configuration.


The thermodynamic results for $16\times 16$ theoretical Model C  
are shown in Figure \ref{eckTD16} 
There is a strong peak in specific heat near ${\cal T}\approx 0.22$ and a more rounded peak 
in $\chi_{xx} \approx \chi_{yy}$ at a slightly higher temperature.  
For all of the models
studied, the out-of-plane magnetic susceptibility $\chi_{zz}$ is considerably smaller than $\chi_{xx}$,
and there is only a weak temperature dependence.  Notably, Model C does reach thermodynamic
equilibrium in the simulations.  This is seen clearly in the plots of the order parameters $Z$ and
$\rho_m$.  Now $\rho_m$, for both discrete and continuous forms, tends towards zero
at low temperature, as expected for the system moving towards a ground state.  Further, the ground
state overlap order parameter, $Z$, tends to go towards unity as $T\rightarrow 0$; this is the
strongest indication of approaching one of the ground states.  It is by chance that the system
ended in $Z=+1$; it could have reached $Z=-1$ with the same probability. At higher temperatures above the
peaks in $C$ and $\chi$ we see that $Z$ becomes quite close to zero; the system is more random and
far from a ground state.  In the same high-temperature region, the monopole density tends towards 
the limiting values, 3/4 for the discrete formula and 7/15 for the continuous definition.  The 
discrete definition for $\rho_m$ necessarily undergoes a stronger change in value as the system makes 
a transition from its low to high temperature behavior.  On the other hand, the larger value
for the continuous definition at low $T$ gives an indication of the fluctuations of island dipoles 
around their long axes as $T\rightarrow 0$.

\section{Hysteresis calculations}
\label{hyster}

\begin{figure}
\includegraphics[width=\smallfigwidth,angle=-90]{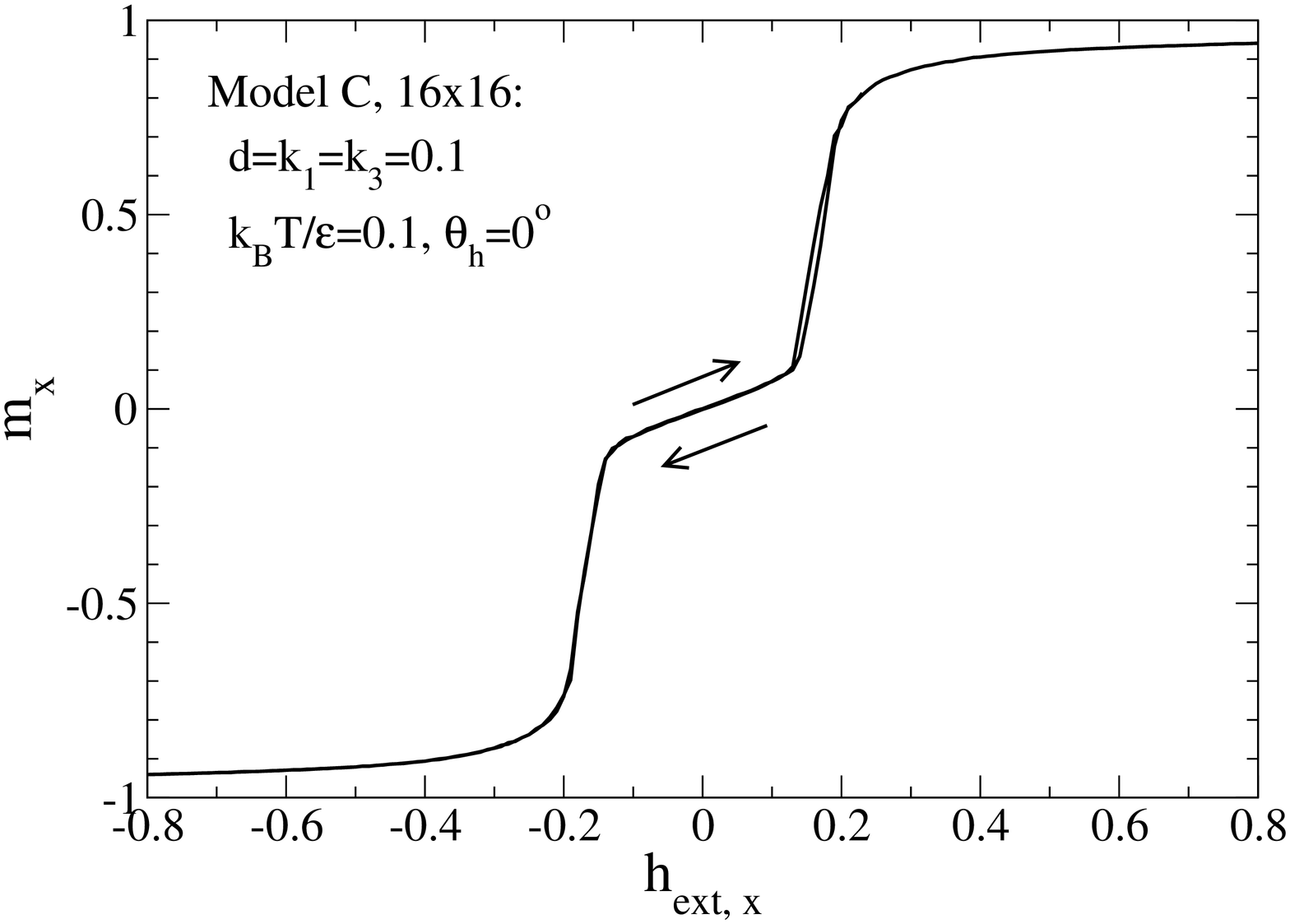}
\includegraphics[width=\smallfigwidth,angle=-90]{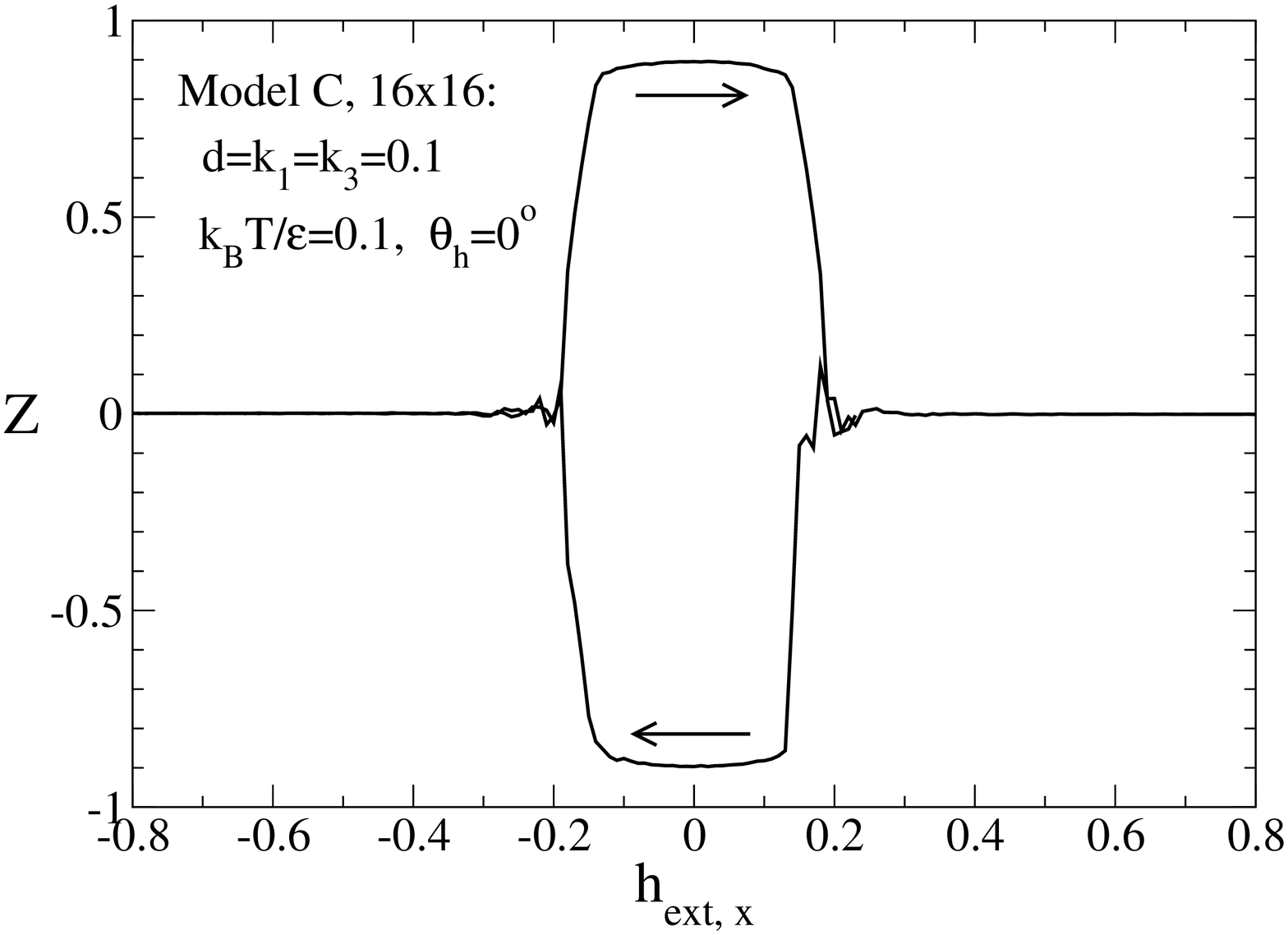}
\caption{\label{D16T4_mh} (Model C) With $D=L_1=K_3=\tfrac{1}{10}\varepsilon$,  at temperature
$k_B T=\tfrac{1}{10}\varepsilon$, (a) the averaged magnetization per site versus external magnetic
field $h_{\rm ext}$ applied along the $x$-axis; (b) the order parameter $Z$ versus $h_{\rm ext}$.
The field strength was initially set at $h_{\rm ext}=0.8$, and scanned to $h_{\rm ext}=-0.8$,
then back to the starting value as in a hysteresis loop calculation.}
\end{figure}

\begin{figure}
\includegraphics[width=\smallfigwidth,angle=-90]{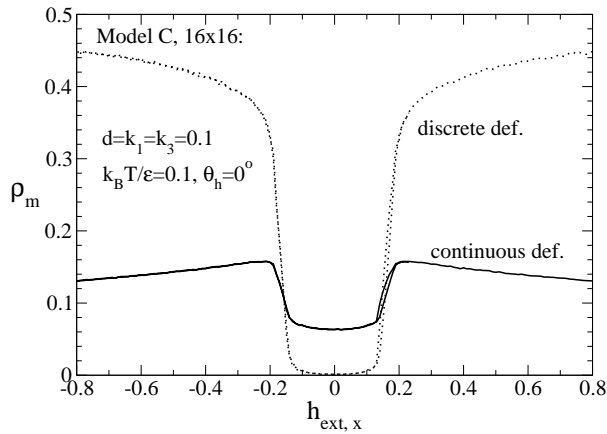}
\caption{\label{D16T4_nqh} (Model C) With $D=K_1=K_3=\tfrac{1}{10}\varepsilon$,  at temperature
$k_B T=\tfrac{1}{10}\varepsilon$, the averaged monopole density versus external magnetic field
$h_{\rm ext}$ applied along the $x$-axis, for the calculations in Fig.\ \ref{D16T4_mh}. The results
of both the discrete and continuous charge definitions are compared here.}
\end{figure}

\begin{figure}
\includegraphics[width=\smallfigwidth,angle=-90]{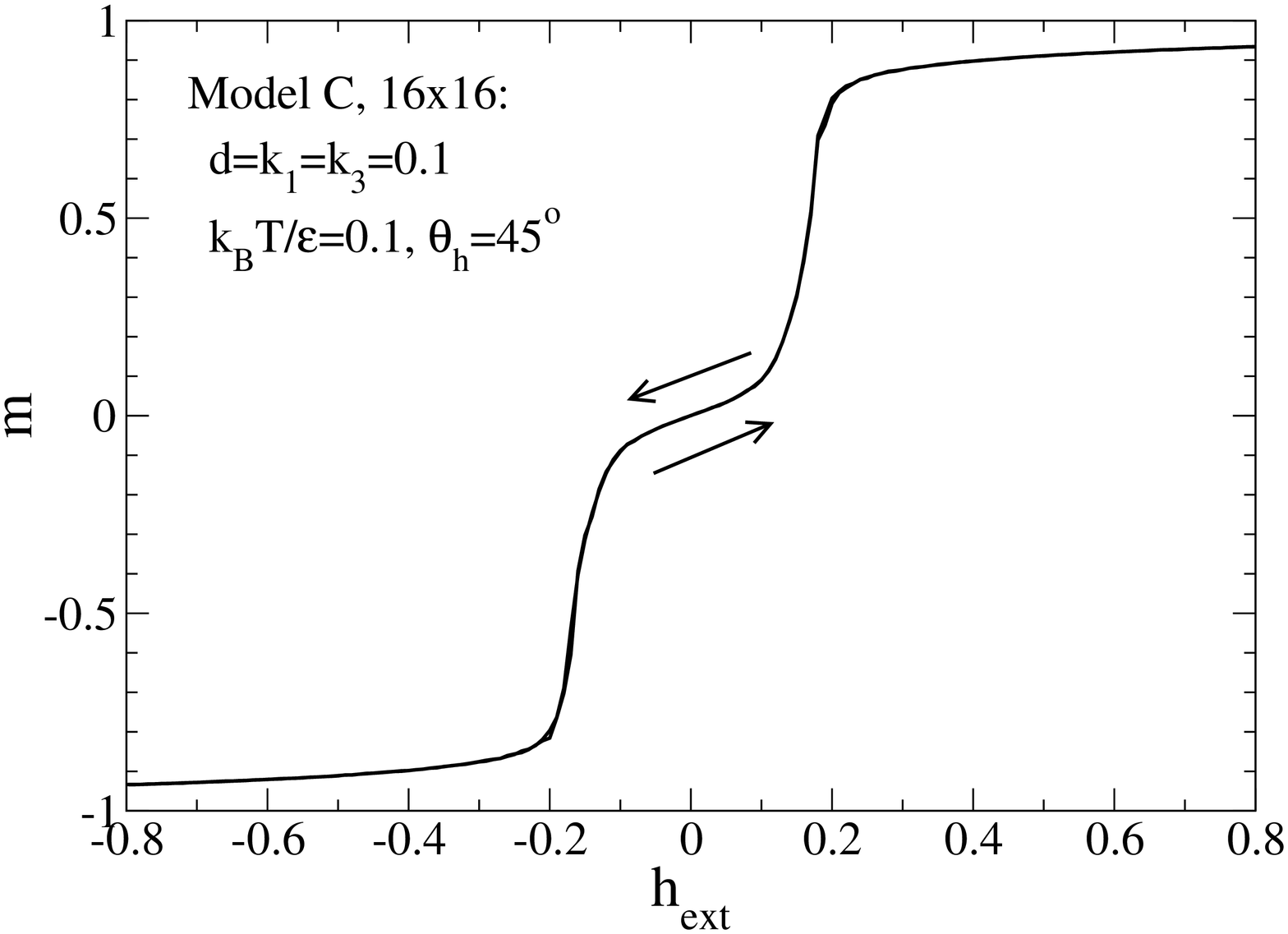}
\includegraphics[width=\smallfigwidth,angle=-90]{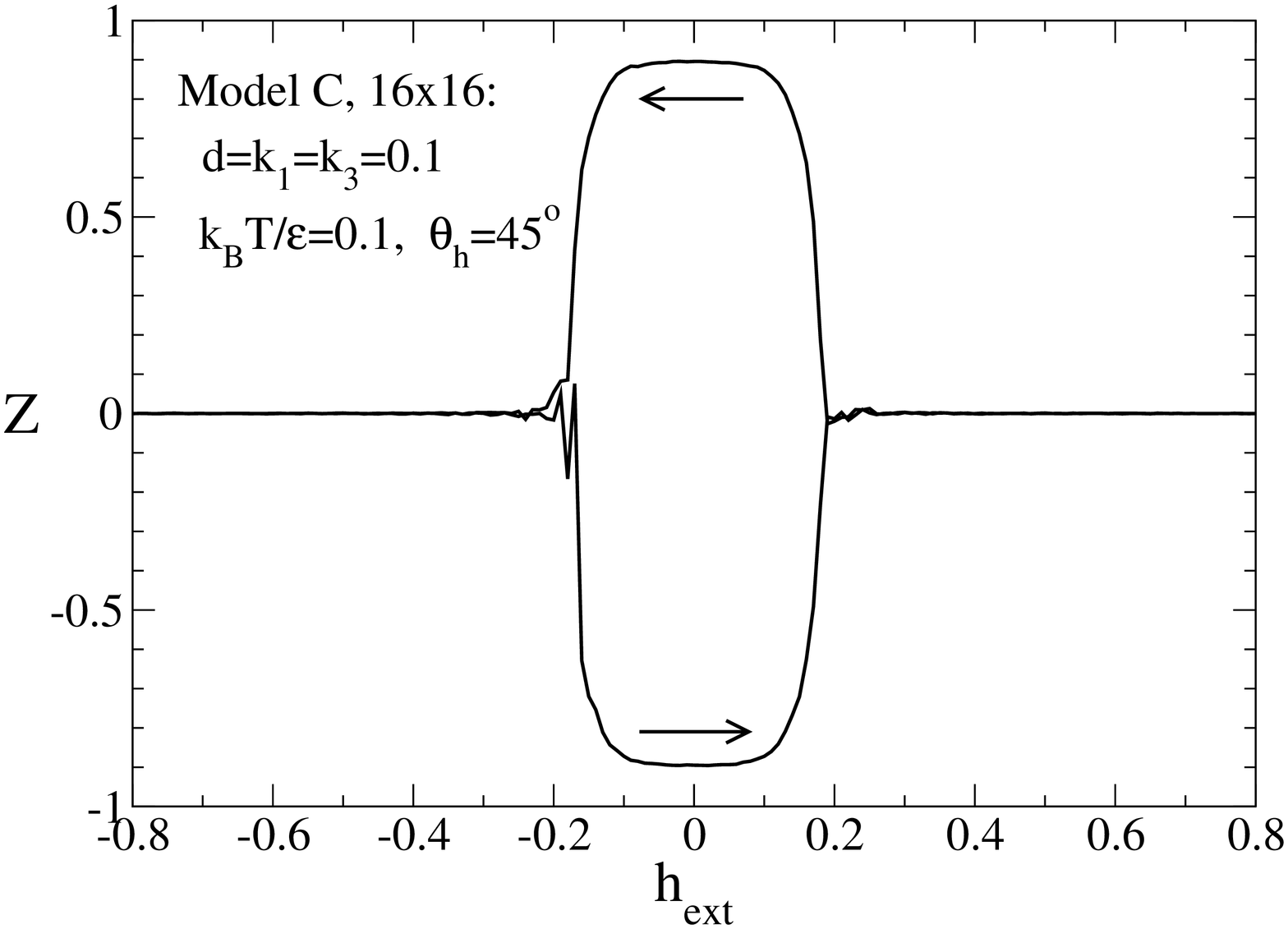}
\caption{\label{45D16T4_mh} (Model C) With $D=K_1=K_3=\tfrac{1}{10}\varepsilon$,  at temperature
$k_B T=\tfrac{1}{10}\varepsilon$, (a) the averaged magnetization per site versus external magnetic
field $h_{\rm ext}$ applied at 45$^{\circ}$ above the $x$-axis; (b) the order parameter $Z$ versus $h_{\rm ext}$.
The field strength was initially set at $h_{\rm ext}=0.8$, and scanned to $h_{\rm ext}=-0.8$,
then back to the starting value as in a hysteresis loop calculation.}
\end{figure}

\begin{figure}
\includegraphics[width=\smallfigwidth,angle=-90]{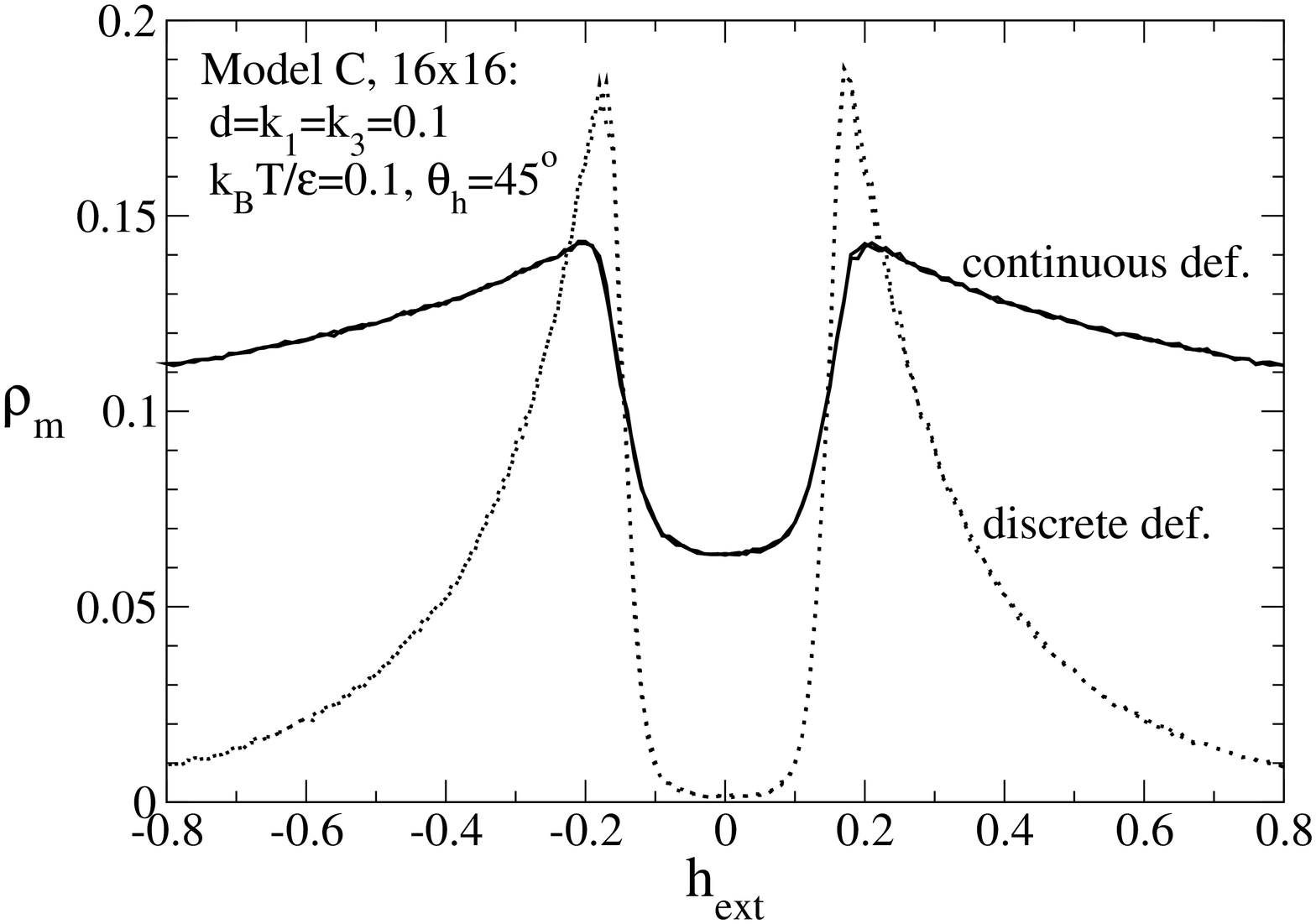}
\caption{\label{45D16T4_nqh} (Model C) With $D=K_1=K_3=\tfrac{1}{10}\varepsilon$,  at temperature
$k_B T=\tfrac{1}{10}\varepsilon$, the averaged monopole density versus external magnetic field
$h_{\rm ext}$ applied at 45$^{\circ}$ above the $x$-axis, for the calculations in Fig.\ \ref{45D16T4_mh}.}
\end{figure}

A simple experiment to investigate the magnetic properties of spin ice is the response in an
applied external field (hysteresis calculation).  To get a general impression of the physical
response in any square spin ice, hysteresis calculations were carried out for Model C at fixed
scaled temperature ${\cal T}=0.1$ .  These were calculated the same as for the thermodynamics, except
that it was adequate to average over shorter sequences, $N_s=1000$, at each step of the applied field.
The system was initially set in a random configuration, but with the maximum positive applied
field.   The field was scanned to lower and negative values along some axis (either $\hat{x}$ or
at 45$^{\circ}$ to $+\hat{x}$) and then allowed to come back to the starting value.  In order
to interpret the results, it was also important to calculate the order parameter $Z$ and the
monopole charge density $\rho_m$ during the hysteresis scan.

The results for field applied along the $\hat{x}$ axis are shown in Figures \ref{D16T4_mh} and
\ref{D16T4_nqh}.  In fact, at this temperature, the model does not exhibit any hysteresis: the
magnetization per island is the same in backward and forward scans of $\vec{h}_{\rm ext}$.
However, the magnetization shows regions with distinctly different slopes.  In Fig.\ \ref{D16T4_mh}b,
one sees that the order parameter $Z$, however, tends to take on either values close to zero,
at strong applied field, or, values near $Z\approx \pm 1$, at weaker field.  We note that this
temperature ${\cal T}=0.1$ is on the low side of the specific heat peak for zero magnetic field.
Then this shows that in the central region of the MH graph, the system falls into states that are close
to the ground states.  These states, however, are slightly modified due to tilting of some
of the dipoles according to the field strength.  Hence, there is close to a linear response with
$h_{\rm ext}$, as the dipoles on the 2$^{\rm nd}$ sublattice, which are nearly perpendicular to
the field, get tilted by it.

By chance, the system in Fig.\ \ref{D16T4_mh} chose $Z=-1$ on the forward scan and $Z=+1$ on the 
reverse scan. These two states are transformed from one into the other simply by reversing the 
choice of the $1$ and $2$ sublattices.  Thus, there is no breaking of this symmetry caused by
the applied field.  There is nothing to prevent both forward and reverse scans from falling
into the same state of $Z$.

The variation of monopole charge density with applied field, for the simulation in Fig.\ \ref{D16T4_mh},
is shown in  Fig.\ \ref{D16T4_nqh}.   Both the discrete and continuous definitions are displayed.
The discrete definition has more dramatic changes.
Especially, $\rho_m$ tends to zero (or a small value for the continuous version) over the same
applied field range where $Z\approx \pm 1$.  This confirms clearly that the central region of the
MH graph corresponds to the system being in states close to the ground states.

For applied field along an axis at 45$^{\circ}$ to the $+\hat{x}$-axis, the situation is similar,
see Figures \ref{45D16T4_mh} and \ref{45D16T4_nqh}.  The MH and ZH graphs are nearly the same as for
those for applied field along $\hat{x}$.  In this case, however, the applied field must be causing
both sublattice dipoles to tilt at stronger fields.  There is a difference, then, in the charge density
plot, see Figure \ref{45D16T4_nqh}.  Again, in the central region near weak $h_{\rm ext}$, the
monopole density tends to zero, as expected for the system being close to one of the ground states.
At strong field, however, the discrete charge density also tends towards zero (as does $Z$).
As the dipoles all tend to align at 45$^{\circ}$ to $\pm\hat{x}$, the net number of dipoles pointing 
into any charge site is then forced to be zero. This clearly forces the monopole density found by the 
discrete definition to zero.   One sees that the monopole density by the continuous definition, 
on the other hand, does not fall to zero at high field, and instead behaves the same as it does 
for field applied along $\hat{x}$.

\section{Discussion and conclusions}
\label{conclude}

We have studied the possibilities of spin dynamics in frustrated artificial spin ice systems 
consisting of two-dimensional square lattices of elongated magnetic nanoislands. The internal 
structure of the magnetic nanoislands was taken into account by assuming quasi-single-domain structure.
Then, depending on the island shapes, aspect ratios, sizes, elements 
and organization in the array, we have looked for possible departures from the usual Ising-like behavior.
We have found that the systems without real dynamics (islands practically with 
an effective Ising behavior) have great difficulty to achieve the ground state (models A \& B). The 
order parameter $Z$ (defined in Section \ref{model}) never reaches the values $-1$ or $+1$ (the two degenerate 
ground states), even for very low temperatures. This result agrees with all experimental studies 
\cite{Wang06,Morgan11} concerning square spin ice.  On the other hand, by considering fictitious material 
constants $D$, $K_{1}$ and $K_{3}$, we found interesting deviations from Ising behavior and consequently, 
more easily thermalized spin dynamics within the array. For this type of system (model C), the ground state 
can be easily obtained for low temperatures. 

Some of the results obtained here can be directly compared to those of Ref.\ \cite{Silva12}, where the 
artificial square spin ice with point-like dipoles with Ising-like behavior was studied by using conventional 
Monte Carlo simulations.  This model with point-like Ising dipoles will be referred to as model I  
(see, for instance, Refs.\ \cite{Mol09,Mol10,Moller09,Silva12}).  Its ground state ($Z = \pm 1$) 
for square lattice ice appears naturally for very low temperatures, i.e., by using conventional Monte 
Carlo simulations.  However, our results for models A \& B, obtained from Langevin dynamics, showed 
that the ground state does not appear at low temperatures.  This may indicate that there is a kind of dynamic 
constraint (effectively, excessively long relaxation time) that prevents it from 
reaching the ground state over a moderate time of observation. Indeed, a similar result is found when the 
dynamics of the model I in the presence of external magnetic fields is 
considered\cite{Budrikis12,Budrikis10,Budrikis11,Budrikis12b,Levis12}. This may indicate that artificial square 
spin ices made with permalloy may never reach the ground state, since both external field dynamics and thermally 
driven dynamics have bottlenecks that prevent access to the ground state. On the other hand, our results for 
model C, for materials with fictitious constants, indicate that it may be possible to access the ground state using 
another kind of material. By changing the island's anisotropies and interactions, the dynamical 
bottleneck can be eliminated.

This difference may be associated with the way that the system explores the phase space. First, consider 
Ising-like islands.  Since we are dealing with classical particles, each 
island must pass through an energy barrier to change its magnetization direction, 
either by the creation and propagation of a domain wall or by rotation of a single domain. Either way, there 
is no option for tunneling and some energy must flow to the island. This internal energy barrier was not taken 
into account in the Monte Carlo calculations of Refs.~\cite{Mol09,Mol10,Silva12} and probably this is why the 
ground state was obtained. The results of the present study and those from 
Refs.~ \cite{Budrikis12,Budrikis10,Budrikis11,Budrikis12b} include the energy barrier for spin flips and we 
may expect that the existence of a huge energy barrier is responsible for the difficulty to access 
the ground state. Moreover, in model C, where the energy barrier is smaller, the ground state is accessible. 
The key factor blocking access to the ground state of artificial square spin ice is the energy 
barrier for spin flips.

For model I, the specific heat exhibits a peak at a temperature around $7.2D/k_{B}$, where it is suspected that 
the string connecting the Nambu monopoles is broken and the system is able to support free monopoles \cite{Silva12}. 
The specific heat for models A, B \& C also exhibits a characteristic peak. For models  A \& B, 
the peak appears for  temperatures around $20D/k_{B}$. 
However, this value is determined more so by the anisotropy $K_1$ and not by $D$. Nonetheless, 
the ground state cannot be obtained for models A \& B even for zero temperature and, therefore, 
there is not a clear way of establishing equivalences between the results of these models and results of model I. 
Even so, the peak in the specific heat for model C occurs at a temperature around $2D/k_{B}$, about three times 
smaller than in model I. This is expected because there are more spin degrees of freedom for model C 
than for model I. Furthermore, the specific heat peak moves to higher temperature  with increasing $K_3$,
as expected from the greater restriction of out-of-plane motion it causes.

To complete this study, we also calculated the hysteresis for the model that exhibits dynamics. It is an 
important calculation to get a general impression of the physical response to an applied external field;
the system tends to pass close to a ground state, as indicated by $Z\approx 0$ near the center of the MH loop. 

The investigations developed here could help 
experimental advances toward spin ice systems in which the ground state could be achieved and/or the transition 
rendered by appearance of free monopoles occurs around room temperature. Experimentally, a recent work was already 
addressed in this direction. Indeed, Kapaklis \textit{et al.}\cite{Kapaklis12} have proposed an experimental 
system (in an external magnetic field) where thermal dynamics can be introduced by varying the temperature of the 
array. On a square lattice, they use a material (based on $\delta$-doped $Pd(Fe)$) with an ordering temperature near 
room temperature to confirm a dynamical ``pre-melting'' of the artificial spin ice structure at a temperature well 
below the intrinsic ordering temperature of the island material. Such a procedure is capable of creating a spin ice 
array that has real thermal dynamics of the artificial spins over an extended temperature range \cite{Kapaklis12}. 
The possibility of observing emergent monopoles is therefore conceivable, following the general approach that the 
authors of Ref.\ \cite{Kapaklis12} described in the design of spin ice arrays. This is a first step towards 
realization of artificial spin ices as conceived in model C, considering some freedom in the selection of its 
parameters.

\section*{Acknowledgments}
The authors thank CNPq, FAPEMIG and CAPES (Brazilian agencies) for financial support; GMW gratefully acknowledges 
the hospitality and support of the Department of Physics at Universidade Federal de Vi\c cosa where this work
was carried out.

\appendix

\section{Langevin dynamics}
The dynamics is investigated here using a Langevin approach \cite{Garcia+98,Nowak00}.  This includes a 
damping term and a rapidly fluctuating stochastic torque in the dynamics.  The size of the stochastic 
torques is related to the temperature and the damping constant, such that the system tends 
towards thermal equilibrium for the chosen temperature.  The approach also gives the dynamics 
at zero temperature but with the damping still included.

In practice, the dynamics is determined by \textit{random magnetic fields}.  This is an approach
considered to be multiplicative noise \cite{Kamppeter+99,Depondt09}, and most importantly, it gives the correct
equilibrium dynamics.  The dynamical equation for some selected unit dipole exposed to a deterministic field
$\vec{h}$ and a stochastic field $\vec{h}_s$ is written in the dimensionless quantities as
\be
\label{LV}
\frac{d{\hat\mu}}{d\tau} = \hat\mu \times \left(\vec{h}+\vec{h}_s \right)
-\alpha \hat\mu \times \left[ (\hat\mu \times \left(\vec{h}+\vec{h}_s \right) \right].
\ee
The first term is the free motion and the second term is the Landau-Gilbert damping, with
dimensionless damping constant $\alpha$.  
For the stochastic fields to establish thermal equilibrium, their time correlations
are determined by the fluctuation-dissipation (FD) theorem,
\be
\langle h_s^i(\tau) \, h_s^j(\tau')\rangle = 2\alpha \,  {\cal T}\, \delta_{ij} \, \delta(\tau-\tau') .
\ee
The indices $i,j$ refer to any of the Cartesian coordinates.
The dimensionless temperature ${\cal T}$ is the thermal energy scaled by the energy unit,
\be
{\cal T} = \frac{k_B T}{\varepsilon} = \frac{k_B T}{\mu\mu_0 M_s}.
\ee
The fluctuation-dissipation theorem indicates that the power in the thermal fluctuations is
carried equivalently in the random magnetic fields.  For reference, in physical units 
the FD relation is
\be
\gamma \mu \langle B_s^i(t) B_s^j(t')\rangle = 2\alpha \, k_B T\, \delta_{ij} \, \delta(t-t') .
\ee

The Langevin equation in (\ref{LV}) is a first-order differential equation where the
noise is multiplicative.  
To discuss the solution method, it is simplest to let $y=y(\tau)$ be a vector that represents the 
entire set of spins, $y=\{\hat\mu_i(\tau)\}$.  Then symbolically $y$ obeys a differential equation  
in the general form,
\be
\label{simple}
\frac{dy}{d\tau} = f[\tau,y(\tau)] + f_s[\tau,y(\tau)] \cdot {h}_s(\tau).
\ee
The function $f$ represents the deterministic time derivative on the RHS
of (\ref{LV}) and the function $f_s$ represents the stochastic part of the dynamics.
Each is defined indirectly by comparing this with the Langevin equation.   The fields $f$,
$f_s$ and $h_s$ are vectors of $3N$ components, where $N$ is the number of dipoles in
the array.

\section{Second order Heun integrator}
An efficient method for integrating this magnetic dynamics type of equation forward in time is the
second order Heun method \cite{Garcia+98,Nowak00}.  That is in the family of predictor-corrector
schemes and is rather stable. 

The predictor stage for the second order Heun algorithm is an Euler step, which is followed by
a corrector stage that is equivalent to the trapezoid rule.   Each involves moving forward in
time over some time step $\Delta \tau$, with the needed results obtained by integrating Eq.\
\ref{simple} from an initial time $\tau_n$ to a final time $\tau_{n+1}=\tau_n+\Delta \tau$,
during which the stochastic fields are acting.  With notation $y_n\equiv y(\tau_n)$, the
predictor stage produces an initial solution estimate $\tilde{y}_{n+1}$ at the end of one time step,
\be
\tilde{y}_{n+1}  = y_n + f(\tau_n,y_n) \Delta \tau + f_s(\tau_n,y_n) \cdot (\sigma_s w_n).
\ee
The effect of the random fields is contained in the last term. The factor $\sigma_s w_n$ replaces
the time-integral of the stochastic magnetic fields. For each site $l$ of the array, there is a
triple of unit variance, zero mean random numbers $(w_{ln}^x, w_{ln}^y, w_{ln}^z)$ produced by a
random number generator.  The physical variance $\sigma_s$ needed in the stochastic fields is
defined by an equilibrium average over the time step. For an individual component at one site,
that is
\bn
\sigma_s &=& \sqrt{\left\langle \left( \int_{\tau_n}^{\tau_{n+1}} d\tau ~ h_s^x(\tau) \right)^2 \right\rangle}
= 
\sqrt{2\alpha {\cal T}\, \Delta \tau}.
\en
Thus, the integrated stochastic field components are replaced by random numbers of zero mean 
with the variance $\sigma_s$.

In the corrector stage, the points $y_n$ and $\tilde{y}_{n+1}$ are used to get better
estimates of the slope of the solution. Their average effect becomes
\bn
y_{n+1} &=& y_n + \frac{1}{2}\left[ f(\tau_n,y_n) +f(\tau_{n+1},\tilde{y}_{n+1}) \right] \Delta\tau
\\
&+& \frac{1}{2} \left[ f_s(\tau_n,y_n) + f_s(\tau_{n+1},\tilde{y}_{n+1}) \right] \cdot (\sigma_s w_n).
\nonumber
\en
It is important to note that the \textit{same} random numbers $w_n$ are used in this corrector
stage as those applied in the predictor stage, for this individual time step.

The change in any spin over a time step, $\Delta \hat\mu = \int d\tau \frac{d}{d\tau} \hat\mu$, 
depends linearly on $\vec{h} \Delta\tau$ (deterministic) and linearly on $\int d\tau ~ h_s(\tau)$ (stochastic).  
The stochastic contribution is replaced by random numbers of the correct variance,
\be
\int_{\tau_n}^{\tau_n+\Delta \tau}  d\tau ~ h_s^x(\tau) \longrightarrow \sigma_s w_n^x.
\ee
Then the Euler predictor step is carried out by evaluating the combined deterministic plus stochastic 
field contributions, for an individual site, like
\bn
\widetilde{\hat\mu} &=& \hat\mu + \Delta \hat\mu,
\\
\Delta \hat\mu &=& \hat\mu \times \left[ \vec{g} -\alpha (\hat\mu\times \vec{g}) \right].
\en
The effective field that updates this site is a combination,
\be
\vec{g} = \vec{h}[\hat\mu]\, \Delta\tau + \sigma_s \vec{w} .
\ee
The same type of combination applies in the trapezoid corrector stage,
The updating field at the end of the time step is calculated
using the \textit{predicted position} together with the \textit{same random fields},
\be
\widetilde{\vec{g}} = \vec{h}[\widetilde{\hat\mu}]\, \Delta\tau + \sigma_s \vec{w}
\ee
That leads to a different estimate for the spin change,
\be
\widetilde{\Delta} \hat\mu = \widetilde{\hat\mu} \times \left[ \widetilde{\vec{g}}
-\alpha (\widetilde{\hat\mu}\times \widetilde{\vec{g}}) \right].
\ee
Then the corrector stage gives the updated spin according to their average
\be
\hat\mu(\tau +\Delta\tau) = \hat\mu(\tau) + \frac{1}{2} \left( \Delta \hat\mu + \widetilde{\Delta} \hat\mu \right).
\ee
This algorithm does not ensure the conservation of spin length.  Thus, the length of $\hat\mu$ can
be rescaled to unity after the step.  

The integration requires a sequence of quasi-random numbers (the $\vec{w}_n$ stochastic fields )
with a long period. 
We have used the generator {\tt mzran13} due to Marsaglia and Zaman \cite{mzran94}, implemented in 
the C-language for long integers. This generator is very simple and fast and has a period of about $2^{125}$. 

\section{Dipole fields on an ice lattice}
The calculation of the dipole term in the local magnetic field, (\ref{hi}), consumes
most of the calculational effort.  
We consider a system with open boundaries.  
There are $N(N-1)/2$ dipole field contributions to be found at any time.  

One of the best ways to speed up the calculation of the dipole fields for larger systems is
to write their calculation as a convolution of a Green's function with the source dipoles,
and calculate that convolution in reciprocal space, transforming between real and reciprocal space
with a fast Fourier transform (FFT) \cite{Sasaki97}.  
We consider that the spin ice involves unit cells on a square lattice,
where each cell has a two-atom basis.  Our approach also would work
for other ice lattices with a different basis.  For the cell whose lower left corner
is at position $\vec{r}_k=(x_k,y_k)=(m_k,n_k)a$, define the two dipoles present.  On the ``1'' sublattice,
\be 
\vec{\mu}_1(\vec{r}_{k}) \text{  at  } \vec{r}_{k1} = (x_k+\tfrac{1}{2}a,y_k),
\ee
and on the ``2'' sublattice,
\be
\vec{\mu}_2(\vec{r}_{k}) \text{  at  } \vec{r}_{k2} = (x_k,y_k+\tfrac{1}{2}a).
\ee
If there is an arbitrary
source dipole $\vec\mu$ at the origin, then the dipolar field $\vec{h}^d$ it creates at position
$\vec{r}=(x,y,z)$, according to the first term in (\ref{hi}), is well-known,
\be
\label{G3x3}
\vec{h}^d(\vec{r}) = \frac{d}{r^5}
\left( \begin{array}{ccc} 2x^2-y^2 & 3xy & 0 \\ 3xy & 2y^2-x^2 & 0 \\ 0 & 0 & -r^2 \end{array} \right)
\cdot
\left( \begin{array}{c} \mu^x \\ \mu^y \\ \mu^z \end{array} \right)
\ee
(with all distances measured in lattice constants).
This can be used to get the field produced from either sublattice.  Summing over source dipoles, the
$3\times 3$ matrix is a Green's operator $\widetilde{G}(\vec{r})$ acting on the dipoles at discrete lattice sites.
(Here the tilde is used only to indicate a $3 \times 3$ matrix quantity.)
However, to account for the two-atom basis, the Green's matrix is expanded to have an extra pair of indices
that refer to the sublattice, one for the field point ($\alpha$) and one for the source point ($\beta$).
The Green's matrix for the field produced at point $\vec{r}_{k\alpha}$ due to the source dipole at
point $\vec{r}_{l\beta}$ is
\begin{widetext}
\be
\widetilde{G}_{\alpha\beta}(\vec{r}_{k},\vec{r}_{l}) \equiv
\frac{d}{\vert \vec{r}_{k\alpha}-\vec{r}_{l\beta}\vert^5}   
\left( \begin{array}{ccc} 2(x_{k\alpha}-x_{l\beta})^2-(y_{k\alpha}-y_{l\beta})^2
& 3(x_{k\alpha}-x_{l\beta})(y_{k\alpha}-y_{l\beta})& 0 \\
3(x_{k\alpha}-x_{l\beta})(y_{k\alpha}-y_{l\beta}) &
2(y_{k\alpha}-y_{l\beta})^2-(x_{k\alpha}-x_{l\beta})^2  & 0 \\ 0 & 0 &
-\vert \vec{r}_{k\alpha}-\vec{r}_{l\beta}\vert^2 \end{array} \right). 
\ee
\end{widetext}
It is important to keep in mind that $\widetilde{G}_{\alpha\beta}$ actually depends
only on the differences of the unit cell positions, $\vec{r}_{kl}\equiv \vec{r}_k-\vec{r}_l$.
Now the dipole field on the $\alpha$ sublattice, for the cell at $\vec{r}_{k}$, is given
by a discrete convolution
\be
\vec{h}^d_{\alpha}(\vec{r}_{k}) = \sum_{l=1}^{N_c} \sum_{\beta=1}^{2}
\widetilde{G}_{\alpha\beta}(\vec{r}_k,\vec{r}_l)\cdot \vec\mu_{\beta}(\vec{r}_l).
\ee
Using fairly obvious notation, $\vec\mu_{\beta}(\vec{r}_l)$ is the dipole on the $\beta$ sublattice
for the unit cell at $\vec{r}_l$.  The dot operation represents the matrix multiplication, i.e., an
implicit sum over the Cartesian components of $\widetilde{G}_{\alpha\beta}$ and $\vec\mu_{\beta}$.
Written this way, the same formula could apply to other lattices of interest, such as honeycomb,
Kagom\'e, etc.  Note that even at $\vec{r}_{k}-\vec{r}_{l}=0$, there are contributions that must
be included, corresponding to the interactions between the sublattices within an individual unit
cell. For a specific example using the square ice sites, one can see that one particular interaction
involving the two different sublattices ($\alpha=1, \beta=2$) has a Cartesian element,
\be
\label{G12}
G^{xx}_{12}(\vec{r}_k,\vec{r}_l) = d \frac{2(x_k-x_l+\frac{a}{2})^2-(y_k-y_l-\frac{a}{2})^2}
{\left[(x_k-x_l+\frac{a}{2})^2+(y_k-y_l-\frac{a}{2})^2\right]^{5/2}}
\ee
This is nonzero when $\vec{r}_k=\vec{r}_l$. Also, the element $G^{xx}_{21}(\vec{r}_k,\vec{r}_l)$
with source and observer sublattices interchanged can be obtained by changing the sign of $a$;
they are not the same. A similar term with source and observer on the {\em same} sublattice is
\be
G^{xx}_{11}(\vec{r}_k,\vec{r}_l) = d \frac{2(x_k-x_l)^2-(y_k-y_l)^2}
{\left[(x_k-x_l)^2+(y_k-y_l)^2\right]^{5/2}}
\ee
This is equal to $G^{xx}_{22}(\vec{r}_k,\vec{r}_l)$.  It is divergent at $\vec{r}_k=\vec{r}_l$,
however, that is a self-interaction that must be excluded by definition.

$\widetilde{G}_{\alpha\beta}$ depends only on the displacements,
$\vec{r}_{kl}\equiv \vec{r}_k-\vec{r}_l$, which form another square lattice.  Then one can find
its Fourier transform, using a fast Fourier transform (FFT), setting the arbitrary source point
to the origin. The Fourier transform of $\vec{\mu}_{\beta}$ is also determined.  The convolution
in real space becomes a simple product of $\widetilde{G}_{\alpha\beta}$ and $\vec{\mu}_{\beta}$
in Fourier space, which can then be transformed back to real space by an inverse FFT to obtain
$\vec{h}^d$. Although there is considerable overhead, for larger systems the speedup is tremendous
($N \ln N$ operations) when compared to doing the $N$ sums with $N$ terms to get the local
dipole fields.

To apply the simplest FFT method, the size of the grid of primitive cells must be
$2^{N_x}\times 2^{N_y}$ with integers $N_x$ and $N_y$.  To avoid the wraparound problem, so that
the system being simulated is really a single copy of the desired $L_x \times L_y$ size, one needs
to choose $N_x$ and $N_y$ large enough so that $2^{N_x}> 2L_x$, and $2^{N_y}> 2L_y$.  This ensures
that the periodic copies of the system, inherent in the application of the Fourier transform,
do not ``see'' or interfere with each other in the convolution.

There are some symmetries that reduce the calculational overhead.  Displacements only on the 
1-sublattice or only on the 2-sublattice are the same, so for any of its Cartesian components,
\be
\widetilde{G}_{11} = \widetilde{G}_{22}\ .
\ee
Also, the matrix is symmetric in the Cartesian indices, for any sublattice indices,
\be
G_{\alpha\beta}^{xy}= G_{\alpha\beta}^{yx}\ .
\ee
Furthermore, the interactions with both source and observer on the same sublattice are
symmetrical in their interchange,
\be
\widetilde{G}_{11}(\vec{r}_{kl})=\widetilde{G}_{11}(\vec{r}_{lk})=
\widetilde{G}_{22}(\vec{r}_{kl})=\widetilde{G}_{22}(\vec{r}_{lk})\ .
\ee
The Fourier transforms of $G_{\alpha\alpha}$ are pure real, leading to some reduction in the
computations needed.
However, there is no symmetry between different sublattices on different unit
cells, so $\widetilde{G}_{12}(\vec{r}_{kl})\ne \widetilde{G}_{21}(\vec{r}_{kl})$,
see the discussion after Eq.\ (\ref{G12}).  The are no self-interactions within a cell, so we do 
define $\widetilde{G}_{11}(0) = \widetilde{G}_{22}(0)=0$.  The interactions between sublattices on the
same cell depend only on squared displacements, so $\widetilde{G}_{12}(0)=\widetilde{G}_{21}(0)\ne 0$.
For $\vec{r}_{kl}\ne0$, these symmetries result in 12 independent elements in
$\widetilde{G}(\vec{r}_{kl})$ (each of $\widetilde{G}_{11}$, $\widetilde{G}_{12}$, and
$\widetilde{G}_{21}$ have 4 independent elements), in contrast to the 4 independent elements
needed for a single sublattice, see the matrix in Eq.\ (\ref{G3x3}).


\section*{References}

\begin{thebibliography}{99}

\bibitem{Anderson56} P.W. Anderson, Phys. Rev. \textbf{102}, 1008 (1956).

\bibitem{Balents10} L. Balents, Nature \textbf{464}, 199 (2010).

\bibitem{Moessner06} R. Moessner and A.R. Ramirez, Phys. Today \textbf{59}, 24 (2006).

\bibitem{Castelnovo08} C. Castelnovo, R. Moessner, and S.L. Sondhi, Nature \textbf{451}, 42 (2008).

\bibitem{Ryzhkin05} I.A. Ryzhkin, JETP \textbf{101}, 481 (2005).

\bibitem{Wang06} R.F. Wang, C. Nisoli, R.S. Freitas, J. Li, W. McConville,
B.J. Cooley, M.S. Lund, N. Samarth, C. Leighton, V.H.  Crespi and P. Schiffer,
Nature {\bf 439}, 303 (2006).

\bibitem{Li10} J. Li, X. Ke, S. Zhang, D. Garand, C. Nisoli P. Lammert, V.H. Crespi, and P. Schiffer,
Phys. Rev. B \textbf{81}, 092406 (2010).

\bibitem{Ladak10} S. Ladak, D.E. Read, G.K. Perkins, L.F. Cohen, and W.R. Brandford,
Nature Phys. \textbf{6}, 359 (2010).

\bibitem{Mengotti11} E. Mengotti, L.J. Heyderman, A.F. Rodriguez, F. Nolting, R.V. H\"{u}gli,
and H-B Braun, Nature Phys. \textbf{7}, 68 (2011).

\bibitem{Mol12} L.A.S. M\'{o}l, A.R. Pereira, and W.A. Moura-Melo, Phys. Rev. B \textbf{85}, 184410 (2012).

\bibitem{Reichhardt12} C. J. Olson Reichhardt, A. Libsal and C. Reichhardt, New
J. Phys. \textbf{14} 025006 (2012).

\bibitem{Budrikis12} Z. Budrikis, K. L. Livesey, J. P. Morgan, J. Akerman, A. Stein, S. Langridge, C. H. Marrows 
and R. L. Stamps, New J. Phys. \textbf{14}, 035014 (2012). 

\bibitem{Silva12b} R. C. Silva, R. J. C. Lopes, L. A. S. Mól, W. A. Moura-Melo, G. M. Wysin and A. R. Pereira
	Phys. Rev. B \textbf{87}, 014414 (2013).

\bibitem{Nascimento12} F. S. Nascimento, L. A. S. M\'{o}l, W. A. Moura-Melo and A. R. Pereira,  New J. Phys. \textbf{14}, 115019 (2012).

\bibitem{Morgan11} J.P. Morgan, A. Stein, S. Langridge, and C. Marrows, Nature Phys. \textbf{7}, 75 (2011).

\bibitem{Mol09} L.A.S. M\'{o}l, R.L. Silva, R.C. Silva, A.R. Pereira, W.A. Moura-Melo,
and B.V. Costa, J. Appl. Phys. \textbf{106}, 063913 (2009).

\bibitem{Mol10} L.A.S. M\'{o}l, W.A. Moura-Melo, and A.R. Pereira, Phys. Rev. B \textbf{82}, 054434 (2010).

\bibitem{Moller09} G. M\"{o}ller and R. Moessner, Phys. Rev. B \textbf{80}, 140409(R) (2009).

\bibitem{Nambu74} Y. Nambu, Phys. Rev. D \textbf{10}, 4262 (1974).

\bibitem{Nisoli10} C. Nisoli, J. Li, X. Ke, D. Garandi, P. Schiffer, and V.H. Crespi,
Phys. Rev. Lett. \textbf{105}, 047205 (2010).

\bibitem{Kapaklis12} V. Kapaklis, U. B. Arnalds, A. Harman-Clarke, E. Th. Papaioannou, M. Karimipour,
P.Korelis, A. Taroni P. C. W. Holdsworth, S. T. Bramwell, and B. Hj\"{o}orvarsson, New J. Phys. \textbf{14},
035009 (2012).

\bibitem{Silva12} R.C. Silva, F.S. Nascimento, L.A. S. M\'{o}l, W.A. Moura-Melo, and A.R. Pereira,
New J. Phys. \textbf{14}, 015008 (2012).


\bibitem{Wysin+12} G.M. Wysin, W.A. Moura-Melo, L.A.S. M\'ol and A.R. Periera,
        J. Phys.: Condens. Matter {\bf 24} 296001 (2012).




\bibitem{Wei03} Zung-Hang Wei, Mei-Feng Lai, Ching-Ray Chang, N.A. Usov,
J.C. Wu and Jun-Yang Lai,  J. Mag. Magn. Mater. {\bf 272-276}, e563 (2004).

\bibitem{Budrikis10} Z. Budrikis, P. Politi, and R.L. Stamps, Phys. Rev. Lett. \textbf{105}, 017201 (2010).

\bibitem{Budrikis11} Z. Budrikis, P. Politi, and R.L. Stamps, Phys. Rev. Lett. \textbf{107}, 217204 (2011).

\bibitem{Budrikis12b} Z. Budrikis, J. P. Morgan, J. Arkeman, A. Stein,  P. Politi, S. Langridge, 
C.H. Marrows, and R.L. Stamps, Phys. Rev. Lett. \textbf{109}, 037203 (2012).

\bibitem{Levis12} D. Levis and L. F. Cugliandolo, EPL (Europhysics Letters) \textbf{97}, 30002 (2012).


\bibitem{Garcia+98} J.L. Garc\'ia-Palacios and F.J. L\'azaro,
        Phys. Rev. B {\bf 58}, 14937 (1998).

\bibitem{Nowak00} U. Nowak, in \textit{Annual Reviews of Computational Physics IX},
        p. 105, edited by D. Stauffer (World Scientific, Singapore, 2000).

\bibitem{Kamppeter+99} T. Kamppeter, F.G. Mertens, E. Moro, A. S\'anchez and A.R. Bishop,
	Phys. Rev. B {\bf 59}, 11349 (1999).

\bibitem{Depondt09} Ph. Depondt and F.G. Mertens, J. Phys.: Condens. Matter {\bf 21}, 336005 (2009).

\bibitem{mzran94} G. Marsaglia and A.  Zaman, Computers in Physics {\bf 8}, No. 1, 117, (1994).

\bibitem{Sasaki97} J. Sasaki and F. Matsubara  {\it J. Phys. Soc. Japan}, \textbf{66}, 2138 (1997).

        

\end{thebibliography}
\end{document}